\documentstyle[12pt,bbox]{article}

\font\sectionfont = cmr12

\makeatletter
\def\@maketitle{\newpage
 \null
 \vskip 2em
 \begin{center}
        {\sectionfont \@title \par}
        \vskip 1.5em
        {\lineskip .5em \begin{tabular}[t]{c}\@author \end{tabular}\par}
        \vskip 1em
        {\large \@date}
 \end{center}
 \par
 \vskip 1.5em}

\def\section{\@startsection {section}{1}{\z@}{-3.5ex plus -1ex minus
 -.2ex}{2.3ex plus .2ex}{\sectionfont}}
\makeatother

\makeatletter
\long\def\@makecaption#1#2{
 \vskip 10pt
  \setbox\@tempboxa\hbox{{\small\bf#1:} \small#2}
   \ifdim \wd\@tempboxa >\hsize {\small\bf#1:} \small#2\par
    \else \hbox to\hsize{\hfil\box\@tempboxa\hfil} \fi}
\makeatother

\font\teneufm=eufm10 \font\seveneufm=eufm7 \font\fiveeufm=eufm5
\newfam\eufmfam
        \textfont\eufmfam=\teneufm
        \scriptfont\eufmfam=\seveneufm
        \scriptscriptfont\eufmfam=\fiveeufm

\def\romanic#1{{\uppercase\expandafter{\romannumeral #1}}}

\begin{document}

\def\lqq{``}
\def\ie{{\it i.e.\ }}
\def\gapp{\, \raisebox{-.5ex}{$\stackrel{>}{\scriptstyle\sim}$}\, }
\def\lapp{\, \raisebox{-.5ex}{$\stackrel{<}{\scriptstyle\sim}$} \, }
\def\half{\textstyle{1\over 2}}
\newcommand{\beq}{\begin{equation}}
\newcommand{\eeq}{\end{equation}}
\newcommand{\beqar}{\begin{eqnarray}}
\newcommand{\eeqar}{\end{eqnarray}}

\phantom{n}
\vskip 4cm

\noindent{\bf HANBURY-BROWN/TWISS INTERFEROMETRY FOR }

\vskip 6.5pt

\noindent{\bf RELATIVISTIC HEAVY-ION COLLISIONS: THEORETICAL ASPECTS}

\vskip26pt

\parindent 1 true in

Ulrich Heinz\footnote{Work supported by BMBF, DFG and GSI.}
\vskip 13pt

Institut f{\"u}r Theoretische Physik, Universit{\"a}t Regensburg,

\indent D-93040 Regensburg, Germany\footnote{Email: 
Ulrich.Heinz@physik.uni-regensburg.de}

\vskip 26pt

\parindent 0.8 true cm



\noindent{\bf 1.~INTRODUCTION}
         
\vskip 13pt

Relativistic heavy-ion collisions at center of mass energies in the 
range of many to many hundreds of GeV per nucleon are performed with 
the ultimate goal of generating a new state of matter, the quark-gluon 
plasma (QGP). To make a QGP, one must heat and compress nuclear matter 
to such high energy densities that the hadrons overlap, creating a 
homogeneous piece of matter consisting only of the stuff usually 
hidden inside hadrons. In the region thus created the quarks and 
gluons, which are usually imprisoned inside the hadrons, become 
``deconfined", i.e. they are able to travel around freely over regions 
which are large compared to the usual confinement length scale of 
about 1 fm (the typical radius of a hadron). Such matter existed in 
and filled all of the volume of the Early Universe during the first 
microsecond of its life but, as far as we know, it has not been 
recreated in the laborarory anywhere since.

How do we know our heavy-ion collision experiment has been successful 
in creating QGP? The answer to this question is surprisingly complex.
The main reason for this is that quarks and gluons cannot travel over 
large distances outside the hot and dense QGP region and thus cannot 
be detected directly. Everything we can measure in the experiment 
provides therefore only indirect knowledge about the hoped-for QGP 
state. This is in particular true for the bulk of secondary particles 
produced in the collision which are hadrons: they can only be formed 
towards the end of the collision process when the reaction zone, by 
expansion into the surrounding vacuum, has cooled down sufficiently to 
allow the quarks and gluons to re-hadronize.

One approach to ``prove" the making of QGP (which I stress must still 
be complemented by other experimental tests to check for consistency, 
see J\"urgen Schukraft's lectures \cite{Schukraft} for examples) 
tries to reach a complete understanding of the space-time structure 
and dynamical state of the reaction zone at the ``freeze-out point" 
where the measured hadrons decouple. One hopes and by now, from the 
experiments performed at the Brookhaven AGS and CERN SPS during the 
last decade, has accumulated a reasonably convincing body of evidence 
that at decoupling the matter has reached a state of local thermal
and perhaps even chemical equilibrium. It also features strong 
longitudinal and transverse collective (hydrodynamic) expansion. While 
some of the longitudinal expansion may be ``primordial", i.e. due to 
incomplete stopping of the two nuclei in the collision region,
all of the transverse expansion must have been generated dynamically
in the reaction. If it is possible to determine the energy density of 
the state at decoupling and simultaneously its expansion velocity, 
then one can try to extrapolate this state backwards in time to a 
point of vanishing transverse expansion and check whether there the 
energy density was above the critical value of about 1 GeV/fm$^3$ 
where one expects the phase transition to the QGP to occur.  

But how does one measure the energy density at decoupling? To measure 
the total energy of the state is not too difficult: one measures the 
momenta of all emitted particles and their masses and adds them up. 
But to compute the energy density we must also know the volume of the 
reaction zone. And there is no known way to measure this volume 
directly. The collision fireball just doesn't hang around long enough 
so that we could shine light on it (or a suitable other species of 
particles) and measure its size by diffraction.

The only known way to obtain indirect experimental information on the 
space-time structure of the particle emitting source created in a
relativistic nuclear collision is through two-particle intensity
(Hanbury-Brown--Twiss (HBT)) interferometry \cite{BGJ90}. The 
goal of this method is to extract the {\em space-time} structure of 
the source from {\em momentum spectra} which are the only measurable 
quantities, making use of the quantum statistical correlations between 
pairs of identical particles. The basic idea, presented in a very 
naive and oversimplified way which to correct I will spend three hours 
of lectures during this Advanced Summer Institute, is as follows: 
Consider a source (for simplicity spherically symmetric) with radius 
$R$, and the emission of a pair of identical particles from point 
$\bbox{x}_1$ in the source with momentum $\bbox{p}_1$ and point 
$\bbox{x}_2$ with momentum $\bbox{p}_2$. If these two points are well 
separated in phase-space, i.e. they satisfy
 \begin{equation}
 \label{uncertain}
   (x_1^i - x_2^i) (p_1^i - p_2^i) \gg 2 \pi \hbar \, ,
   \qquad i=1 \ or\ i=2\ or\ i=3\,,
 \end{equation}
this process can be treated classically. If on the other hand, 
 \begin{equation}
 \label{uncertain1}
   (x_1^i - x_2^i) (p_1^i - p_2^i) \leq 2 \pi \hbar \, ,
   \qquad i=1,\ 2 \ and\ 3\,,
 \end{equation}
the two particles sit close in phase-space, and quantum mechanics can 
no longer be ignored. The most important quantum mechanical correction 
to be taken into account is the (anti-)symmetrization of the 
two-particle wave function: it ensures vanishing probability for two 
identical fermions to originate from the same phase-space point, and 
for bosons it leads to an enhanced probability to find them at the 
same point in phase-space compared to the classical expectation 
(bosons are ``social subjects").

Since the distance in coordinate space $(x_1^i - x_2^i)$ is limited by 
the finite diameter $2R$ of the source, we can force the system into 
the quantum domain by measuring particle pairs with smaller and 
smaller relative momentum $\bbox{p}_1 - \bbox{p_2}$. Once 
$q_i \equiv p_1^i - p_2^i$ becomes smaller than $\pi\hbar/R$, the two 
particles can no longer avoid quantum mechanics by escaping to larger 
relative distances in coordinate space. There emission probability 
will be affected by wave function symmetrization, leading in the case 
of bosons (fermions) to an enhanced (reduced) pair emission 
probability compared to the classical expectation (which would be 
simply the product of the individual single-particle emission 
probabilities). The two-particle {\em correlation function} is thus 
expected to begin to appreciably deviate from unity for relative 
momenta $q < q^* \simeq (\hbar/R)$. The critical value $q^*$ at which 
this effect sets in (conveniently one chooses for $q^*$ the value 
where the correlation is half way between it maximum (minimum) value 
and 1) is thus a measure for the geometric radius $R$ of the source.

So far the naive picture of how two-particle HBT interferometry works.
Unfortunately, the only situation where it applies more or less 
directly is for photon interferometry of stars for which the method 
was invented. I will spend my time here in Dronten to explain to you 
why this naive picture is generically {\em wrong} and how to 
substitute it correctly. The basic reason why the above simple cartoon 
ceases to work in high energy nuclear and particle physics is that
the sources created in hadronic or heavy-ion collisions live only for 
very short time periods and feature inhomogeneous temperature 
profiles and strong collective dynamical expansion. I will show you 
that for such sources the HBT radius parameters (half widths of the 
correlation function) generally don't measure the full source size, 
but only so-called ``space-time regions of homogeneity" inside which 
the momentum distribution varies sufficiently little so that the 
particles can actually show the quantum statistical correlations. 
They also mix the spatial and temporal structure of the source, and we 
will learn tricks (in particular a new way of parametrizing
the correlation functions) which are designed unfold these different 
aspects from the data. 

The size of the just mentioned homogeneity regions varies with the 
momentum of the emitted particles, causing an important dependence of 
the HBT parameters on the pair momentum. I will show you how this 
momentum dependence can be used to extract the strength of the 
collective flow of the source at decoupling. To do so in a 
quantitative way requires the detailed consideration of many physical 
features of the particle emission process. One such feature which I
will discuss here in some detail is the fact that 
often some of the particles one uses in constructing the 
pair-correlation function don't come directly from the source but are 
created well after decoupling by the decay of unstable resonances.  
Resonance decays not only affect the size and momentum dependence of 
some of the HBT radius parameters but also the overall strength of the 
correlation as described by the so-called ``chaoticity parameter" 
$\lambda$.  

My lectures are structured in the following way: In the First 
Lecture, I discuss the general connection of the measured one- and 
two-particle spectra with the phase-space distribution of particles in 
the source. In particular I discuss the important aspects of ``chaotic" 
versus ``coherent" particle emission and how to implement them in a 
formal way. This establishes the formal background which I exploit in 
the Second Lecture to extract in a more quantitative way general 
(``model-independent") analytic relationships between the geometric 
and dynamic space-time structure of the source and certain features of 
the pair correlation function. In the Third Lecture I analyze these 
relationships quantitatively in the framework of a general class of 
model sources with finite geometric extension and three-dimensional
collective expansion. The results should give you a good qualitative 
and even semi-quantitative feeling of the expected behaviour of the 
pair correlation function in relativistic heavy-ion collisions.

The Fourth Lecture should have presented a comparison of the 
calculations with the data, but fortunately the organizers gave me 
only three hours time so that I was spared the embarrassment of having 
to admit that at the present moment no quantitative such comparison
exists: the theoretical analysis is still so new that most of the 
diagrams you will see are less than a few months old, and the new 
high-quality heavy-ion data which allow for the detailed 
multidimensional analysis advocated here are still so hot 
(``preliminary") that the experimentalists wouldn't give me permission 
to show them anyhow. Please look into Barbara Jacak's lecture notes in 
this volume \cite{Jacak} and into the Proceedings of the {\em Quark 
Matter '96} Conference in Heidelberg, 20-24 May 1996, if you want to 
catch a first glimpse of the data.  

\vskip 26pt

\noindent{\bf 2.~LECTURE 1: SPECTRA AND EMISSION FUNCTION} 

\vskip 13pt


\noindent{\bf 2.1.~One- and two-particle spectra} 

\vskip 13pt

The covariant single- and two-particle distributions are defined by
 \begin{eqnarray}
   P_1(\bbox{p}) 
  & = & E\, \frac{dN}{d^3p} 
        = E \, \langle\hat{a}^+_{\bbox{p}} \hat{a}_{\bbox{p}}\rangle \, ,
 \label{1} \\
   P_2(\bbox{p}_a,\bbox{p}_b) 
  & = & E_a\, E_b\, \frac{dN}{d^3p_a d^3p_b}
        = E_a \, E_b\, 
          \langle\hat{a}^+_{\bbox{p}_a} \hat{a}^+_{\bbox{p}_b}
                 \hat{a}_{\bbox{p}_b} \hat{a}_{\bbox{p}_a} \rangle \, ,
 \label{2}
 \end{eqnarray}
where $\hat{a}^+_{\bbox{p}}$ ($\hat{a}_{\bbox{p}}$) creates (destroys) a
particle with momentum $\bbox{p}$. The angular brackets denote an 
ensemble average,
 \begin{equation}
 \label{aver}
    \langle \hat O \rangle = {\rm tr}\, (\hat \rho \hat O)\, ,
 \end{equation} 
where $\hat \rho$ is the density operator associated with the 
ensemble. (When talking about an ensemble we may think of either a 
single large, thermalized source, or a large number of similar, but 
not necessarily thermalized collision events.) The single-particle 
spectrum is normalized to the average number of particles, $\langle N 
\rangle$, per collision, 
 \begin{equation}
 \label{norm1}
   \int {d^3p \over E}\, P_1(\bbox{p}) = \langle N \rangle \, ,
 \end{equation}
while the two-particle distribution is normalized to the number 
of possible pairs, $\langle N (N-1) \rangle$, per event:
 \begin{equation}
 \label{norm2}
   \int {d^3p_a \over E_a}\,{d^3p_b \over E_b}\, P_2(\bbox{p}_a,\bbox{p}_b) 
   = \langle N (N-1) \rangle \, .
 \end{equation}
The two-particle correlation function is defined as \cite{gyul}
 \begin{equation}
 \label{3}
   C(\bbox{p}_a,\bbox{p}_b) 
   = \frac{\langle N\rangle^2}{\langle N(N-1)\rangle}\,
     \frac{P_2(\bbox{p}_a,\bbox{p}_b)}{P_1(\bbox{p}_a)P_1(\bbox{p}_b)} \, .
 \end{equation}
If the two particles are emitted independently and final state 
interactions are neglected I will show that it is possible to prove a
generalized Wick theorem,
 \begin{equation}
   P_2(\bbox{p}_a,\bbox{p}_b) 
   = \frac{\langle N(N-1)\rangle}{\langle N\rangle^2}\,
     \Bigl( P_1(\bbox{p}_a)P_1(\bbox{p}_b) 
            \pm |\bar{S}(\bbox{p}_a,\bbox{p}_b)|^2 \Bigr) \;,
 \label{3a}
 \end{equation}
where the plus (minus) sign refers to bosons (fermions), and we have 
defined the following covariant quantity:
 \begin{equation}
   \bar{S}(\bbox{p}_a,\bbox{p}_b) = \sqrt{E_a E_b}\,
   \langle \hat{a}^+_{\bbox{p}_a} \hat{a}_{\bbox{p}_b} \rangle\;.
 \label{5}
 \end{equation}
If the ensemble corresponds to a thermalized source in global 
thermodynamic equilibrium, i.e. $\hat \rho$ is the (grand) canonical 
density operator, this is just the well-known ``thermal Wick theorem"
\cite{FW}. But even in nonthermal sources emission may be ``chaotic",
i.e. the emission of particle $a$ at point $x_a$ may be completely 
independent from the emission of particle $b$ at $x_b$. This can be
described by assigning the wave function of the emitted particle
a random phase $\phi$ which is averaged over in the ensemble average. 
I will show that this is also sufficient to guarantee Wick's theorem
(\ref{5}). -- The opposite to independent, ``chaotic" particle 
emission is emission from a coherent state (e.g. in a laser) where the 
phases of all emitted particles are fully correlated with each other; 
in that case the second term in (\ref{5}) is completely missing. The 
general situation can be parametrized by a superposition of density 
operators, 
 \begin{equation}
 \label{rho}
  \hat \rho = \alpha \, \hat \rho_{\rm chaotic} 
            + (1-\alpha)\, \hat \rho_{\rm coherent} \,  ,
 \end{equation}
with $0\leq \alpha \leq 1$, where only the chaotic first part 
contributes to the second term in (\ref{3a}): 
 \begin{eqnarray}
 \label{rho1}
  \langle \hat a^+_a \hat a^+_b \hat a_b \hat a_a \rangle
  &{=}& \langle \hat a^+_a \hat a_a \rangle 
      \langle \hat a^+_b \hat a_b \rangle 
      \pm \alpha \vert \langle \hat a^+_a \hat a_b \rangle_{\rm ch} \vert^2
 \nonumber\\
  &{+}& \alpha(1-\alpha) \left[
  \left( \langle \hat a^+_a \hat a_a \rangle_{\rm ch} 
       - \langle \hat a^+_a \hat a_a \rangle_{\rm coh} \right)
  \left( \langle \hat a^+_b \hat a_b \rangle_{\rm ch} 
       - \langle \hat a^+_b \hat a_b \rangle_{\rm coh} \right)
  \right] \, .    
 \end{eqnarray}   
The last term contributes only if the chaotic and coherent parts of 
the density operator generate different single-particle spectra. One 
easily checks that for $\alpha <1$ in such an ensemble the 
correlations are ``incomplete", i.e. at $\bbox{p}_a=\bbox{p}_b$ the 
correlation function (\ref{3}) approaches a value less than 2 for 
bosons and larger than 0 for fermions. -- Since so far the heavy ion 
data don't appear to require a coherent component in the particle 
production process, I will for the remainder of these lectures assume 
$\alpha=1$ and refer the interested reader for the possible effects 
from partial coherence to the literature \cite{marburg}.  

Assuming the generalized Wick theorem (\ref{3a}) the correlation 
function (\ref{3}) can be written as
 \begin{equation}
 \label{corr}
  C(\bbox{p}_a, \bbox{p}_b) = 1 \pm 
  {\vert \langle \hat a^+_{\bbox{p}_a} \hat a_{\bbox{p}_b} \rangle \vert^2
   \over
   \langle \hat a^+_{\bbox{p}_a} \hat a_{\bbox{p}_a} \rangle 
   \langle \hat a^+_{\bbox{p}_b} \hat a_{\bbox{p}_b} \rangle } \, .
 \end{equation}
Note that the second term is positive definite, i.e. the correlation 
function cannot, for example, oscillate around unity. [If you see such 
a behaviour in the literature \cite{sriv} (and the authors did not 
include final state interactions) it is wrong.]

From now on I will assume that the emitted particles are bosons, and 
for convenience I will call them pions, although nearly everything in 
the first two lectures applies equally well to other bosonic particles 
(except where explicitly stated). In the last Lecture I will be a 
little more specific and distinguish between 2-pion and 2-kaon 
correlations for comparison.

\vskip 13pt

\noindent{\bf 2.2.~The generalized Wick theorem} 

\vskip 13pt

Pions are created in heavy-ion collisions throughout the history of 
the collision process, but we are only interested in those pions which 
reach the detector as free, non-interacting particles. (Unfortunately,
pion interferometry can only be practically performed with charged pions, 
because neutral pions decay too rapidly. But charged pions have a 
long-range final state interaction, the Coulomb repulsion from
the other pion in the pair with equal charge, as well as the attractive 
or repulsive Coulomb interaction with the hundreds of other produced 
charged particles, including the charge of the protons in the 
fireball. A proper treatment of this many-body Coulomb problem is a 
difficult task \cite{BBM}. Since I don't know yet how to do it 
properly, I will simply neglect the Coulomb interaction in the final 
state -- assuming that somehow the experimentalists know how to 
approximately correct their measured correlations for it 
\cite{Jacak}.) 

We assume that the last interaction of the pion, in which the finally 
observed pion is created in its free asymptotic state, can be
parametrized by a classical source amplitude $J(x)$. The solution
of the free Klein-Gordon equation for pions generated by such a 
classical current,
 \begin{equation}
 \label{KG}
   \Bigl( \Box + m^2 \Bigr) \hat \phi(x) = J(x) \, ,
 \end{equation}
with outgoing boundary conditions is given \cite{gyul} by a classical 
(``coherent") state
 \begin{equation}
 \label{coherent}
  \vert J\rangle = e^{-{\bar n}/2} \, \exp\left( i\int d^3p \, 
  \tilde J(\bbox{p})\, \hat a^+_{\bbox{p}} \right) \vert 0 \rangle
 \end{equation}
where
 \begin{equation}
   \tilde{J}(\bbox{p}) = \int \frac{d^4x}{\sqrt{(2\pi)^3 2 E_p}}\,
   \exp[i(E_p t-\bbox{p}{\cdot}\bbox{x})] \, J(x)
 \label{8}
 \end{equation}
is the on-shell Fourier transform of the source $J(x)$, and the 
normalization of the state is given by 
 \begin{equation}
   \bar n = \int d^3p\, \vert \tilde{J}(\bbox{p})\vert^2 \, .
 \label{8a}
 \end{equation}
The state (\ref{coherent}) is an eigenstate of the destruction operator:
 \begin{equation}
   \hat{a}_{\bbox{p}} \vert J \rangle = i\tilde{J}(\bbox{p})|J\rangle \, .
 \label{8b}
 \end{equation}

\vskip 13pt

\noindent{\it 2.2.1.~Emission from a single coherent state} 

\vskip 13pt

If there is only a single classical source $J(x)$, the corresponding 
density operator of the ``ensemble" is just the projection operator on 
the coherent state, $\hat \rho_{\rm coherent} = \vert J \rangle 
\langle J \vert$, and, using (\ref{8b}), the single- and two-particle 
spectra (\ref{1},\ref{2}) are easily evaluated:
 \begin{eqnarray}
 \label{coh1}
  E {dN\over d^3 p} &=& E\, 
  \langle J\vert \hat a^+_{\bbox{p}} \hat a_{\bbox{p}} \vert J \rangle =  
  E \vert \tilde J(\bbox{p}) \vert^2\, ,
 \\
 \label{coh2}
  E_a E_b {dN \over d^3p_a \, d^3p_b} &=& E_a E_b \,
  \langle J\vert \hat a^+_{\bbox{p}_a} \hat a^+_{\bbox{p}_b} 
                 \hat a_{\bbox{p}_b} \hat a_{\bbox{p}_a} \vert J \rangle =  
  E_a \vert \tilde J(\bbox{p}_a) \vert^2 \cdot 
  E_b \vert \tilde J(\bbox{p}_b) \vert^2\, .
 \end{eqnarray}
Obviously, there is no exchange term in the two-particle spectrum 
which is simply given by the product of the single-particle spectra. 
{\em A coherent state thus has no Bose-Einstein correlations.}

\vskip 13pt

\noindent{\it 2.2.2.~Emission by a chaotic superposition of 
classical sources} 

\vskip 13pt

This changes if we consider a superposition of classical source 
amplitudes each of which emits free pions independently, i.e. with a 
random phase \cite{gyul}:
 \begin{equation}
    J(x) 
    = \sum_{i=1}^N e^{i\phi_i}\,e^{-ip_i{\cdot}(x-x_i)}\,
      J_0(x-x_i)\, .
 \label{Jx}
 \end{equation}
The construction rule \cite{kole,CH94} for this source is obvious: we 
take $N$ sources $J_0(x)$ with identical internal structure, give each 
of them a boost with 4-momentum $p_i$, then translate them to 
different positions $x_i$ in the fireball and supply them with a 
random phase $\phi_i$. This allows for arbitrary $x$-$p$ correlations 
\cite{kole} (i.e. correlations between the momentum spectrum of the 
emitted particles and the point from where they are emitted). The 
momenta $p_i$ of the sources can, but need not be on the pion 
mass-shell; for example, the source could be a decaying 
$\Delta$-resonance with 3-momentum $\bbox{p}_i$. The on-shell Fourier 
transform of (\ref{Jx}) is 
 \begin{equation}
    \tilde{J}(\bbox{p}) 
    = \sum_{i=1}^N e^{i\phi_i}\, e^{ip{\cdot}x_i}\, 
      \tilde{J}_0(p-p_i)\, ,
 \label{onshell}
 \end{equation}
where
 \begin{equation}
   \tilde{J}_0(p-p_i) = \int \frac{d^4x}{\sqrt{(2\pi)^3 2 E_p}}\,
   e^{i(p-p_i){\cdot}x}\, J_0(x)
 \label{J0}
 \end{equation}
is the (regular) Fourier transform of $J_0(x)$, and $p$ is on-shell 
while $p_i$ may be off-shell. The state $\vert J \rangle$ which is 
defined by inserting (\ref{onshell}) into (\ref{coherent}) now depends 
on the parameters $\{ x_i,p_i,\phi_i;i=1,\dots,N\}$:
 \begin{equation}
 \label{state}
   \vert J \rangle \equiv \big\vert J[N;\{x,p,\phi\}]\big\rangle\, .
 \end{equation}
The ensemble of sources can be defined in terms of a density operator 
$\hat \rho$ which fixes the distribution of these parameters. We 
assume that the number of sources $N$ is distributed with a 
probability distribution $P_N$, the phases $\phi$ are distributed 
randomly between 0 and $2\pi$, and the source positions $x_i$ and 
momenta $p_i$ are distributed with a phase-space density $\rho(x,p)$,
with normalizations
 \begin{equation}
    \sum_{N=0}^\infty P_N = 1 \, ,\qquad
    \sum_{N=0}^\infty N\, P_N = \langle N \rangle \, ,\qquad
    \int d^4x\,d^4p\,\rho(x,p) = 1\, .
 \label{7b}
 \end{equation}
The corresponding ensemble average is given by
 \begin{equation}
   {\rm tr}(\hat{\rho}\,\hat O)= \sum_{N=0}^\infty P_N
   \prod_{i=1}^N\int d^4x_i\, d^4p_i\,\rho(x_i,p_i)
   \int_0^{2\pi}\frac{d\phi_i}{2\pi}\,
   \big\langle J[N;\{x,p,\phi\}] \big\vert \hat O \big\vert 
               J[N;\{x,p,\phi\}] \big\rangle \, .
 \label{8c} 
 \end{equation}
The calculation of the single-particle spectrum is straightforward:
 \begin{eqnarray}
   \langle \hat a^+_{\bbox{p}} \hat a_{\bbox{p}} \rangle &=&
   \sum_{N=0}^\infty P_N
   \prod_{i=1}^N\int d^4x_i\, d^4p_i\,\rho(x_i,p_i)
   \int_0^{2\pi}\frac{d\phi_i}{2\pi}
 \nonumber\\
   &\times& \sum_{n,n'=1}^N e^{i(\phi_n-\phi_{n'})}\,
   e^{ip{\cdot}(x_n-x_{n'})}\, \tilde J^*_0(p-p_{n'}) \tilde J_0(p-p_n) \, .
 \label{single}
 \end{eqnarray}
After performing the integrations over the phases $\phi_i$ in the 
double sum over $n$ and $n'$, only the diagonal terms with $n$=$n'$ 
survive. For each term in the remaining single sum over $n$ the 
integrations over $x_i$ and $p_i$, $i\ne n$, can be done using the 
normalization condition (\ref{7b}). After suitably relabelling the 
dummy integration variables for the one remaining $x$- and 
$p$-integration we end up with $N$ identical terms under the sum over 
$n$. This allows to perform the sum over $N$, and we simply get 
 \begin{eqnarray}
   P_1(\bbox{p}) = E_p\,\langle|\tilde{J}(\bbox{p})|^2\rangle
   &=& \langle N\rangle\, E_p\int d^4x' \, d^4p'\,
       \rho(x',p')\, |\tilde{J}_0(p-p')|^2 
 \label{single1}\\
   &=& \langle N\rangle\, E_p\int d^4p'\,
       \tilde{\rho}(p')\,|\tilde{J}_0(p-p')|^2 \;.
 \label{8d}
 \end{eqnarray}
The single particle spectrum is thus obtained by folding the intrinsic 
momentum spectrum $\vert \tilde{J}_0(p)\vert^2$ of the individual 
source currents $J_0$ with the 4-momentum distribution of the sources, 
$\tilde{\rho}(p) = \int d^4x\, \rho(x,p)$. 

The algebra for the two-particle spectrum is a little more involved.
It is useful to first compute 
 \begin{eqnarray}
   \langle \hat a^+_{\bbox{p}_a} \hat a_{\bbox{p}_b} \rangle &=&
   \sum_{N=0}^\infty P_N
   \prod_{i=1}^N\int d^4x_i\, d^4p_i\,\rho(x_i,p_i)
   \int_0^{2\pi}\frac{d\phi_i}{2\pi}
 \nonumber\\
   &\times& \sum_{n,n'=1}^N e^{i(\phi_n-\phi_{n'})}\,
   e^{i(p_b{\cdot}x_n-p_a{\cdot}x_{n'})}\, 
   \tilde J^*_0(p_a-p_{n'}) \tilde J_0(p_b-p_n) \, .
 \label{exch}
 \end{eqnarray}
Again only the terms $n$=$n'$ survive the phase average, and after
doing the dummy integrations over $x_i,p_i$, $i\ne n$, one finds that 
the remaining sum over $n$ contains again $N$ identical terms, such 
that the sum over $N$ can be performed:
 \begin{equation}
   \langle \hat a^+_{\bbox{p}_a} \hat a_{\bbox{p}_b} \rangle 
   = \langle N\rangle\, \int d^4x' \, d^4p'\, \rho(x',p')\, 
     e^{i(p_b-p_a){\cdot}x'}\, \tilde J^*_0(p_a-p') \tilde J_0(p_b-p') \, .
 \label{exch1}
 \end{equation}
With this auxiliary result at hand we can now attack the two-particle 
spectrum. From the definitions one finds
 \begin{eqnarray} 
   \langle \hat a^+_{\bbox{p}_a} \hat a^+_{\bbox{p}_b} 
           \hat a_{\bbox{p}_b} \hat a_{\bbox{p}_a} \rangle &=&
   \sum_{N=0}^\infty P_N
   \prod_{i=1}^N\int d^4x_i\, d^4p_i\,\rho(x_i,p_i)
   \int_0^{2\pi}\frac{d\phi_i}{2\pi}
 \nonumber\\
   &\times& \sum_{n,n',m,m'=1}^N 
   e^{i(\phi_n+\phi_m-\phi_{n'}-\phi_{m'})}\,
   e^{ip_a{\cdot}(x_n-x_{n'})}\, e^{ip_b{\cdot}(x_m-x_{m'})}\, 
 \nonumber\\
   && \quad\times \quad \tilde J^*_0(p_a-p_{n'}) \tilde J^*_0(p_b-p_{m'}) 
   \tilde J_0(p_b-p_m) \tilde J_0(p_a-p_n) \, .
 \label{double}
 \end{eqnarray}
The integration over the phases $\phi_i$ now yields two types of 
nonvanishing contributions: $n=n',m=m'$ and $n=m',m=n'$. The term
where all four summation indices are equal, $n=m=n'=m'$, should be 
omitted \cite{gyul}: it corresponds to emission of both particles from 
the same elementary source, and if one carefully first puts the whole 
system in a finite volume $V$, performs the calculation there and lets 
$V\to\infty$ in the end, then this term is suppressed relative to the 
others by a factor $1/V$. We thus get
 \begin{eqnarray}
   \langle \hat a^+_{\bbox{p}_a} \hat a^+_{\bbox{p}_b} 
           \hat a_{\bbox{p}_b} \hat a_{\bbox{p}_a} \rangle 
   &=& \sum_{N=0}^\infty P_N
   \prod_{i=1}^N\int d^4x_i\, d^4p_i\,\rho(x_i,p_i)
   \sum_{n\ne m}^N \Bigl[ 
   \vert \tilde J_0(p_a-p_n) \vert^2 \,
   \vert \tilde J_0(p_b-p_m) \vert^2 
 \nonumber\\
   &+& e^{i(p_a-p_b){\cdot}(x_n-x_m)}\, 
   \tilde J^*_0(p_a-p_m) \tilde J^*_0(p_b-p_n) 
   \tilde J_0(p_b-p_m) \tilde J_0(p_a-p_n) \Bigr]
 \nonumber\\
   &=& \sum_{N=0}^\infty P_N
   \prod_{i=1}^N\int d^4x_i\, d^4p_i\,\rho(x_i,p_i)
 \nonumber\\
   &\times& \left[ 
    \left(\sum_{n\ne m}^N \vert \tilde J_0(p_a-p_n) \vert^2 \right)
    \left(\sum_{m=1}^N \vert \tilde J_0(p_b-p_m) \vert^2 \right)
    \right.
 \nonumber\\
   && + \left(\sum_{n\ne m}^N e^{i(p_a-p_b){\cdot}x_n}\, 
        \tilde J^*_0(p_b-p_n) \tilde J_0(p_a-p_n) \right)
 \nonumber\\
   && \times \left.
       \left(\sum_{m=1}^N e^{-i(p_a-p_b){\cdot}x_m}\,
       \tilde J_0(p_b-p_m) \tilde J^*_0(p_a-p_m) \right)
      \right]\, .
 \label{double1}
 \end{eqnarray}
After again doing the dummy integrations over $x_i,p_i$, $i \ne n,m$,
one realizes that each of the two terms in the square bracket contains
$N(N-1)$ identical terms, yielding a factor $\langle N(N-1) \rangle$ 
after performing the sum over $N$. Up to this factor, the first term 
is just a product of two terms of the type (\ref{single1}), i.e. a 
product of single-particle spectra, while the second term is
recognized as a product of (\ref{exch1}) and its complex conjugate. 
We thus have
 \begin{eqnarray}
    P_2(\bbox{p}_a,\bbox{p}_b) 
    &=& \frac{\langle N(N-1)\rangle}{\langle N\rangle^2}\, E_a\, E_b 
        \Bigl[ 
        \langle \hat a^+_{\bbox{p}_a} \hat a_{\bbox{p}_a} \rangle
        \langle \hat a^+_{\bbox{p}_b} \hat a_{\bbox{p}_b} \rangle +
        \big\vert
        \langle \hat a^+_{\bbox{p}_a} \hat a_{\bbox{p}_b} \rangle 
        \big\vert^2
        \Bigr]
 \label{Wick}\\
    &=& \frac{\langle N(N-1)\rangle}{\langle N\rangle^2}\, E_a\, E_b 
        \Bigl[ 
        \langle| \tilde{J}({\bf p}_a)|^2\rangle
        \langle| \tilde{J}({\bf p}_b)|^2\rangle +
        \big\vert
        \langle \tilde{J}^*({\bf p}_a) \tilde{J}({\bf p}_b)\rangle
        \big\vert^2
      \Bigr]\, ,
 \label{Wick1}
 \end{eqnarray}
which proves the generalized Wick theorem (\ref{3a}).

\vskip 13pt

\noindent{\bf 2.3.~Source Wigner function and spectra} 

\vskip 13pt

These expressions can be rewritten in a very nice and suggestive way 
by introducing the so-called ``emission function" $S(x,K)$ 
\cite{shuryak,pratt,marburg}:
 \begin{equation}
   S(x,K) = \int\frac{d^4y}{2(2\pi)^3}\, e^{-iK{\cdot}y}
   \left\langle J^*(x+\half y)J(x-\half y)\right\rangle \, .
 \label{8f}
 \end{equation}
It is the Wigner transform of the density matrix associated with the 
classical source amplitudes $J(x)$. This Wigner density is a quantum 
mechanical object defined in phase-space $(x,K)$; in general it is 
neither positive definite nor real. But, when integrated over $x$ or 
$K$ it yields the classical (positive definite and real) source 
density in momentum or coordinate space, respectively, in exactly the 
same way as a classical phase-space density would behave. Furthermore, 
textbooks on Wigner functions show that their non-reality and 
non-positivity are genuine quantum effects resulting from the 
uncertainty relation and are concentrated at short phase-space 
distances; when the Wigner function is averaged over phase-space 
volumes which are large compared to the volume $(2\pi\hbar)^3$ of an 
elementary phase-space cell, the result is real and positive definite 
and behaves exactly like a classical phase-space density. 

The emission function $S(x,K)$ is thus the quantum mechanical analogue
of the classical phase-space distribution which gives the probability
of finding at point $x$ a source which emits free pions with momentum 
$K$. Please note that $K$ in $S(x,K)$ can be off-shell. Also, it is 
defined in terms of a 4-dimensional Wigner transform of the source 
density matrix \cite{shuryak}, in contrast to the 3-dimensional 
expression suggested by Pratt \cite{pratt} which neglects retardation 
and off-shell effects.  

Using Eq.~(\ref{8}) it is easy to establish the following relationship:
 \begin{eqnarray}
   \tilde{J}^*(\bbox{p}_a)\, \tilde{J}(\bbox{p}_b) 
   & = &
   \int \frac{d^4x_1\, d^4x_2}{(2\pi)^3 \,2\sqrt{E_aE_b}}\,
   \exp(-ip_a{\cdot}x_1 +ip_b{\cdot}x_2)J^*(x_1)J(x_2)
 \nonumber \\
   & = & 
   \int\frac{d^4x\,d^4y}{(2\pi)^3\,2\sqrt{E_aE_b}}\,
   \exp(-iq{\cdot}x-iK{\cdot}y)J^*(x+\half y)J(x-\half y)  \, ,
 \label{8e}
 \end{eqnarray}
where $x=\half(x_1x_2)$ and $y=x_1-x_2$. Inserting this into 
Eqs.~(\ref{single1}) and (\ref{Wick1}) one finds the fundamental 
relations:
 \begin{eqnarray}
  E_K {dN \over d^3K} &=&
  \int d^4x\, S(x,K) \, ,
 \label{spectrum}\\
  C(\bbox{q},\bbox{K}) &=& 1 + 
  {\left\vert \int d^4x\, S(x,K)\, e^{iq{\cdot}x} \right\vert^2
   \over
   \int d^4x\, S(x,K+\half q) \ \int d^4x\, S(x,K-\half q)}\, .
 \label{correlator}
 \end{eqnarray}
For the single-particle spectrum (\ref{spectrum}), the Wigner function
$S(x,K)$ on the r.h.s. must be evaluated on-shell, i.e. at $K^0=E_K
= \sqrt{m^2 + \bbox{K}^2}$. For the correlator (\ref{correlator}) we 
have defined the relative momentum $\bbox{q} = \bbox{p}_a - 
\bbox{p}_b$, $q^0 = E_a-E_b$ between the two particles in the pair, 
and the total momentum of the pair $\bbox{K} = (\bbox{p}_a + 
\bbox{p}_b)/2$, $K^0= (E_a+E_b)/2$. Of course, since the 4-momenta
$p_{a,b}$ of the two measured particles are on-shell, $p^0_i = E_i =
\sqrt{m^2 + \bbox{p}_i^2}$, the 4-momenta $q$ and $K$ are in general 
off-shell. They satisfy the orthogonality relation
 \begin{equation}
 \label{ortho}
   q \cdot K = 0\,.
 \end{equation}
Thus, the Wigner function on the r.h.s. of Eq.~(\ref{correlator})
is {\em not} evaluated at the on-shell point $K^0 = E_K$. This implies
that for the correlator, in principle, we need to know the off-shell 
behaviour of the emission function, i.e. the quantum mechanical 
structure of the source. Obviously, this makes the problem appear 
rather untractable!  

Fortunately, nature is nice to us: the interesting behaviour of the
correlator (its deviation from unity) is concentrated at small 
values of $\vert \bbox{q} \vert$. Expanding $K^0 = (E_a+E_b)/2$ for 
small $q$ one finds 
 \begin{equation}
 \label{Konshell}
   K^0 = E_K \, \left( 1 + {\bbox{q}^2 \over 8 E_K^2} + 
   {\cal O}\left({\bbox{q}^4 \over E_K^4}\right) \right) 
   \approx E_K \, .
 \end{equation}
Since the relevant range of $q$ is given by the inverse size of
the source (more properly: the inverse size of the regions of 
homogeneity in the source -- see Lecture 2), the validity of this 
approximation is ensured in practice as long as the Compton wavelength 
of the particles is small compared to this ``source size". For the 
case of pion, kaon, or proton interferometry for heavy-ion collisions 
this is true automatically due to the rest mass of the particles: even 
for pions at rest, the Compton wavelength of 1.4 fm is comfortably 
smaller than any typical nuclear source size. This is of enormous 
practical importance because it allows you essentially to replace the 
source Wigner density by a classical phase-space distribution function
for on-shell particles. This provides a necessary theoretical 
foundation for the calculation of HBT correlations from classical 
hydrodynamic or kinetic (e.g. cascade or molecular dynamics) 
simulations of the collision.  

In photon interferometry there is no rest mass available to help you: 
for photons, the approximation $K^0 \approx E_K$ can only be justified 
if they escape from the source with high momentum, and in HBT 
interferometry with soft photons the quantum mechanical nature of the 
emission function needs to be explicitly considered. In practice this 
means that one must study the photon production processes 
microscopically and quantum mechanically. 

If the single-particle spectrum is an exponential function of the 
energy then it is easy to prove \cite{CSH95b} that one can replace the 
product of single-particle distributions in the denominator of 
(\ref{correlator}) by the square of the single-particle spectrum 
evaluated at the average momentum $K$:
 \begin{equation}
 \label{corrapp}
  C(\bbox{q},\bbox{K}) \approx 1 + 
  \left\vert {\int d^4x\, e^{iq{\cdot}x}\, S(x,K) 
              \over
              \int d^4x\, S(x,K)} 
  \right\vert^2
  \equiv 1 + \left\vert \langle e^{iq{\cdot}x} \rangle \right\vert^2
  \, .
 \end{equation}
The deviations from this approximation are proportional to the 
curvature of the single-particle distribution in logarithmic 
representation \cite{CSH95b}. They are small in practice because the 
measured single-particle spectra are usually more or less exponential.
In the second equality of (\ref{corrapp}) we defined $\langle \dots 
\rangle$ as the average taken with the emission function; due to the 
$K$-dependence of $S(x,K)$ this average is a function of $K$.
This notation will be used extensively in Lecture 2.

The ensemble average on the r.h.s. of (\ref{8f}) is defined in the 
sense of Eq.~(\ref{8c}) and can be evaluated with the help of the 
definition (\ref{Jx}). One finds 
 \begin{equation}
   S(x,K) = \langle N \rangle \int d^4z\, d^4q \,\rho(x-z,q) \, 
   S_0(z,K-q) \, ,
 \label{S}
 \end{equation}
where 
 \begin{equation}
   S_0(x,p) = \int\frac{d^4y}{2(2\pi)^3}\, e^{-ip{\cdot}y}
              J_0^*(x+\half y)J_0(x-\half y)  
 \label{S0}
 \end{equation}
is the Wigner function associated with an individual source $J_0$. 
This establishes a similar folding relation for the Wigner function 
itself as we have already obtained in (\ref{8d}) for the
single-particle spectrum: the emission function of the complete source
is obtained by folding the Wigner function for an individual pion
source $J_0$ with the Wigner distribution $\rho$ of these sources. 
Eq.~(\ref{S}) is useful for the calculation of quantum statistical 
correlations from classical Monte Carlo event generators for heavy-ion 
collisions: $\langle N \rangle \rho(x,p)$ can be considered as the 
distribution of the classical phase-space coordinates of the pion 
emitters (decaying resonances or 2-body collision systems), and 
$S_0(x,p)$ as the Wigner function of the free pions emitted at these 
points (for example, a Gaussian in Quantum Molecular Dynamics 
calculations \cite{aichelin}). Replacing the former by a sum of 
$\delta$-functions describing the space-time locations of the last 
interactions and the pion momenta just afterwards, and the latter by 
a product of two Gaussians with momentum spread $\Delta p$ and 
coordinate spread $\Delta x$ such that $\Delta x\Delta p \geq 
\hbar/2$, we recover the expressions derived in \cite{padula}.  

The fundamental relations (\ref{spectrum}) and (\ref{correlator}) 
resp. (\ref{corrapp}) show that {\em both the single-particle spectrum 
and the two-particle correlation function can be expressed as simple 
integrals over the emission function}. The emission function thus is 
the crucial ingredient in the theory of HBT interferometry: if it is 
known, the calculation of one- and two-particle spectra is 
straightforward (even if the evaluation of the integrals may in some 
cases be technically involved); more interestingly, measurements of 
the one- and two-particle spectra provide access to the emission 
function and thus to the space-time structure of the source. This 
latter aspect is, of course, the motivation for exploiting HBT in 
practice. In my second and third Lecture I will concentrate on the 
question to what extent this access to the space-time structure from 
only momentum-space data really works, whether it is complete, and 
(since we will find it is not and HBT analyses will thus be 
necessarily model-dependent) what can be reliably said about the 
extension and dynamical space-time structure of the source anyhow, 
based on a minimal set of intuitive and highly suggestive model 
assumptions.  

%
%
%

\vskip 26pt

\noindent{\bf 3.~LECTURE 2: MODEL-INDEPENDENT DISCUSSION OF HBT
\newline 
\phantom{3.}~CORRELATION FUNCTIONS}

\vskip 13pt

In this lecture I will discuss very general relations between the 
space-time structure of the source (as encoded in the $x$-dependence 
of the emission function $S(x,K)$) and the shape of the two-particle 
correlation function. These relations are valid for arbitrary emission 
functions, and in this sense the discussion is {\it 
model-independent}. It nevertheless provides important insight into 
the physical features of HBT interferometry, in particular for 
short-lived dynamical sources, and it clarifies what HBT can achieve 
and what not.  

\vskip 13pt

\noindent{\bf 3.1.~The mass-shell constraint}

\vskip 13pt

Expressions (\ref{correlator},\ref{corrapp}) show that the correlation 
function is related to the emission function by a Fourier 
transformation. At first sight this might suggest that one should 
easily be able to reconstruct the emission function from the measured 
correlation function by inverse Fourier transformation, the single 
particle spectrum (\ref{spectrum}) providing the normalization. This 
is, however, not correct. The reason is that, since the correlation 
function is constructed from the on-shell momenta of the measured 
particle pairs, not all four components of the relative momentum $q$ 
occurring on the r.h.s. of (\ref{corrapp}) are independent. They are 
related by the ``mass-shell constraint" (\ref{ortho}) which can, for 
instance, be solved for $q^0$:
 \begin{equation}
 \label{massshell}
   q^0 = \bbox{\beta}\cdot \bbox{q} \qquad {\rm with} \qquad 
   \bbox{\beta} = {\bbox{K}\over K^0} \approx {\bbox{K}\over E_K}\, .
 \end{equation} 
$\bbox{\beta}$ is (approximately) the velocity of the c.m. of the 
particle pair. The Fourier transform in (\ref{corrapp}) is therefore 
not invertible, and the reconstruction of the space-time structure of 
the source from HBT measurements will thus always require additional 
model assumptions. 

It is instructive to insert (\ref{massshell}) into (\ref{corrapp}): 
 \begin{equation}
 \label{corrapp1}
  C(\bbox{q},\bbox{K}) \approx 1 + 
  \left\vert {\int d^4x\, \exp\bigl( 
                          i\bbox{q}{\cdot}(\bbox{x}-\bbox{\beta}\, t) 
                          \bigr) \, S(x,K)
              \over
              \int d^4x\, S(x,K)} 
  \right\vert^2 \, .
 \end{equation}
This shows that the correlator $C(\bbox{q},\bbox{K})$ actually mixes the 
spatial and temporal information on the source in a non-trivial way 
which depends on the pair velocity $\bbox{\beta}$. Only for 
time-independent sources things seem to be simple: the correlator then 
just measures the Fourier transform of the spatial source 
distribution. Closer inspection shows, however, that it does so only 
in the directions {\em perpendicular} to $\bbox{\beta}$ since the time 
integration leads to a $\delta$-function 
$\delta(\bbox{\beta}{\cdot}\bbox{q})$: 
 \begin{equation}
 \label{static}
  \lim_{T\to \infty} \left\vert 
  {\int_{-T}^T dt\, \exp\left(-i\,\bbox{q}{\cdot}\bbox{\beta}\, t\right) 
   \over 
   \int_{-T}^T dt} \right\vert^2 =  
  \lim_{T\to \infty} {2\pi \over T} \,
  \delta(\bbox{q}{\cdot}\bbox{\beta}) \, .
 \end{equation}
This implies that there are no correlations in the direction {\em 
parallel} to the pair velocity $\bbox{\beta}$ (which will be called 
the ``outward" direction below), i.e. $C=1$ for $q_{\rm out}\ne 0$. 
The width of the correlator in this direction vanishes! This should 
puzzle you: wouldn't you have thought that the width of the correlator 
in the ``outward" direction is inversely related to the source size in 
that direction (which is, of course, perfectly finite)? As we will 
see in the next subsection this unexpected behaviour is just another 
consequence of the mixing of the spatial and temporal structure of the 
source in the correlator: The width parameter of the correlator in the 
``outward" direction receives also a contribution from the lifetime of 
the source which in this case diverges, leading to the vanishing width 
of the correlator.  

\vskip 13pt

\noindent{\bf 3.2.~$K$-dependence of the correlator}

\vskip 13pt

Eq.~(\ref{corrapp}) shows that in general the correlator is a function 
of {\em both} $\bbox{q}$ and $\bbox{K}$. Only if the emission function 
factorizes in $x$ and $K$, $S(x,K) = F(x)\,G(K)$, which means that every 
point $x$ in the source emits particles with the same momentum 
spectrum $G(K)$ (no ``$x$-$K$-correlations"), the $K$-dependence in 
$G(K)$ cancels between numerator and denominator of (\ref{corrapp}), 
and the correlator seems to be $K$-independent. However, not even 
this is really true: even after the cancellation of the explicit 
$K$-dependence $G(K)$, there remains an implicit $K$-dependence
via the pair velocity $\bbox{\beta} \approx \bbox{K}/E_K$ 
in the exponent on the r.h.s. of Eq.~(\ref{corrapp1})! Only if both 
conditions, factorization of the emission function in $x$ and $K$ {\em 
and} time-independence of the source, apply simultaneously, the 
correlation function is truely $K$-independent (because then the 
$\bbox{\beta}$-dependence resides only in the $\delta$-function 
(\ref{static})). 

The only practical situation which I know where this occurs and a 
$K$-independent correlation function should thus be expected is in HBT 
interferometry of stars for which the method was invented \cite{HBT}.  
It is hard to believe that this complication in the application of the 
original HBT idea to high-energy collisions went nearly unnoticed for 
more than 20 years and was stumbled upon more or less empirically by 
Scott Pratt in his pioneering work on HBT interferometry for heavy-ion 
collisions \cite{pratt} only in 1984!  

If one parametrises it by a Gaussian in $q$ (see below) this means 
that in general the parameters (``HBT radii'') depend on $K$. Typical
sources of $x$-$K$ correlations in the emission function are a
collective expansion of the emitter and/or temperature gradients in 
the particle source: in both cases the momentum spectrum $\sim \exp[-
p{\cdot}u(x)/T(x)]$ of the emitted particles (where $u^\mu(x)$ is the 
4-velocity of the expansion flow) depends on the emission point. In 
the case of collective expansion, the spectra from different emission 
points are Doppler shifted relative to each other. If there are
temperature gradients, e.g. a high temperature in the center and 
cooler matter at the edges, the source will look smaller for 
high-momentum particles (which come mostly from the hot center) than 
for low-momentum ones (which receive larger contributions also from 
the cooler outward regions).

We thus see that collective expansion of the source induces a 
$K$-dependence of the correlation function. But so do temperature 
gradients. The crucial question is: does a careful measurement of the 
correlation function, in particular of its $K$-dependence, permit a 
separation of such effects, i.e. can the collective dynamics of the 
source be quantitatively determined through HBT experiments? We will 
see that this is not an easy task; however, with sufficiently good 
data, it should be possible. In any case, the $K$-dependence of the 
correlator is a decisive feature which puts the HBT game into a 
completely new ball park. Even if it sounds exaggerated and may at 
first offend some of my experimentalist friends who are busy fighting 
the limited statistics of their data: two-particle correlation 
measurements which are not able to resolve the $K$-dependence of the 
HBT parameters are, in high energy nuclear and particle physics, 
essentially useless. [Unfortunately, this applies to all the HBT data 
from $pp$ and $e^+e^-$ collisions which I am aware of. In my opinion, 
a renewed investigation of two-particle correlations from $pp$ 
and $e^+e^-$ collisions, using the powerful new tool of 
multidimensional HBT analysis, should be a high priority project -- as 
it is, we have practically nothing with which to compare our heavy-ion 
results in a meaningful way.] 

\vskip 13pt

\noindent{\bf 3.3.~The Gaussian approximation}

\vskip 13pt

As motivated in the Introduction, the most interesting feature of the 
two-particle correlation function is its half-width. Actually, since 
the relative momentum $\bbox{q} = \bbox{p}_1 - \bbox{p_2}$ has three 
Cartesian components, the fall-off of the correlator for increasing 
$q$ is not described by a single half-width, but rather by a 
(symmetric) 3$\times$3 tensor \cite{CNH95} which describes the 
curvature of the correlation function near $\bbox{q} = 0$. We will see 
that in fact nearly all relevant information that can be extracted 
from the correlation function resides in the 6 independent components 
of this tensor \cite{WSH96}. This in turn implies that in order to 
compute the correlation function $C$ it is sufficient to approximate 
the source function $S$ by a Gaussian in $x$ which contains only 
information on its space-time moments up to second order. [Gaussian 
approximations for the emission function have been used for the 
discussion of HBT correlation functions in many different variants 
\cite{CSH95b,CNH95,WSH96,CSH95a,AS95,CL96,HTWW96,WHTW96}, a perfect 
example how research proceeds by trial and error. Here I give the 
rigorous derivation first published in \cite{HTWW96}.]  

Let us write the arbitrary emission function $S(x,K)$ in
the following form:
 \begin{equation}
 \label{7}
   S(x,K) = N(K)\  S(\bar x(K),K)\ 
            e^{ - \half \tilde x^\mu(K)\, B_{\mu\nu}(K)\, \tilde x^\nu(K)}
   + \delta S(x,K) \, ,
 \end{equation} 
where we adjust the parameters $N(K)$, $\bar x^\mu(K)$, and 
$B_{\mu\nu}(K)$ of the Gaussian first term in such a way that the 
correction term $\delta S$ has vanishing zeroth, first and second 
order space-time moments:
 \begin{equation}
 \label{deltaS}
   \int d^4x\, \delta S(x,K) = 
   \int d^4x\, x^\mu\, \delta S(x,K) = 
   \int d^4x\, x^\mu x^\nu\,  \delta S(x,K) = 0\, .
 \end{equation} 
This is achieved by choosing
 \begin{eqnarray}
  N(K) &=& E_K {dN\over d^3 K}\,
           {\det B_{\mu\nu}(K) \over S(\bar x(K),K)}\, ,
 \label{NK}\\
  \bar x^\mu(K) &=& \langle x^\mu \rangle\, , 
 \label{barx}\\
  \left(B^{-1}\right)_{\mu\nu}(K) 
  &=& \langle \tilde x_\mu \tilde x_\nu \rangle 
      \equiv \langle (x -\bar x)_\mu (x- \bar x)_\nu \rangle \, .
 \label{Bmunu}
 \end{eqnarray}
The ($K$-dependent) average over the source function $\langle \dots 
\rangle$ has been defined in Eq.~(\ref{corrapp}). The normalization factor 
(\ref{NK}) ensures that the Gaussian term in (\ref{7}) gives the 
correct single-particle spectrum (\ref{spectrum}); it fixes the 
normalization on-shell, i.e. for $K^0=E_K$, but as we discussed this 
is where we need the emission function also for the computation of the 
correlator. (Note that for photon interferometry this may not be true, 
and (\ref{NK}) should then be replaced by a suitable generalization.) 
$\bar x(K)$ in (\ref{barx}) is the centre of the emission function 
$S(x,K)$ and approximately equal to its ``saddle point", i.e. the 
point of highest emissivity for particles with momentum $K$. The 
second equality in (\ref{Bmunu}) defines $\tilde x$ as the space-time 
coordinate relative to the centre of the emission function; only this 
quantity enters the further discussion, since, due to the invariance 
of the momentum spectra under arbitrary translations of the source in 
coordinate space, the absolute position of the emission point is not 
measurable in experiments which determine only particle momenta. Since 
$\bar x(K)$ is not measurable, neither is the normalization $N(K)$ 
\cite{HTWW96} as its definition (\ref{NK}) involves the emission 
function at $\bar x(K)$. Finally, Eq.~(\ref{Bmunu}) ensures that the 
Gaussian first term in (\ref{7}) correctly reproduces the variances 
$\langle \tilde x_\mu \tilde x_\nu \rangle$ of the original emission 
function, in particular its r.m.s. widths in the various space-time 
directions.  

Inserting the decomposition (\ref{7}) into Eq.~(\ref{corrapp}) we 
obtain for the correlation function
 \begin{equation}
 \label{corrgauss}
   C(\bbox{q},\bbox{K}) = 1 + \exp\bigl[
   - q^\mu\, q^\nu\, \langle \tilde x_\mu \tilde x_\nu \rangle(\bbox{K}) 
   \bigr]
   + \delta C(\bbox{q},\bbox{K})\, .
 \end{equation}
The Gaussian in $q$ results from the Fourier transform of the Gaussian 
contribution in (\ref{7}); the last term $\delta C$ receives 
contributions from the second term $\delta S$ in (\ref{7}) which 
contains information on the third and higher order space-time moments 
of the emission function, like sharp edges, wiggles, secondary peaks, 
etc. in the source. It is at least of fourth order in $q$, i.e. the 
second derivative of the full correlator at $q=0$ is given {\em 
exactly} by the Gaussian in (\ref{corrgauss}). Please note that the 
exponent of the correlator contains no term linear in $q$; since the 
correlator must be symmetric under $\bbox{q} \to - \bbox{q}$ because 
it does not matter which of the two particles of the pair receives the 
label 1 or 2, a linear $q$-dependence could only arise in the form 
$\exp(-R \vert \bbox{q} \vert)$. The only type of emission function 
yielding such a $q$-dependence of the correlator would be a 
spherically symmetric Lorentzian. Any emission function which at large 
$x$ falls off faster than $1/x^2$ results in the leading Gaussian 
behaviour (\ref{corrgauss}) instead. This settles, in my opinion, the 
old issue whether Gaussian or exponential fits of the correlation 
function should be preferred.  

[In the past it has repeatedly been observed that the correlation data 
appear to be better fit by exponentials than by Gaussians. However, 
as far as I know, this happened always when one tried to fit the 
correlator as a function of the single Lorentz invariant variable 
$Q_{\rm inv}^2 = (q^0)^2 - \bbox{q}^2$. Contemplating the structure
of Eq.~(\ref{corrgauss}) one realizes that such a fit does not make 
sense: the generic structure of the exponent, $-q^\mu q^\nu \langle 
\tilde x_\mu \tilde x_\nu \rangle$, tells us that the term $(q^0)^2$ 
should come with the time variance of the source while the spatial 
components $(q^i)^2$ should come with the spatial variances of the 
source. Since all variances are positive semidefinite by definition, 
it does not make sense to parametrize the correlation function by a 
variable in which $(q^0)^2$ and $\bbox{q}^2$ appear with the opposite 
sign! Such a fit could only work if the time variance and all mixed
variances would vanish identically, and all three spatial variances 
were equal. This is certainly not the general case in nature. The
good exponential fits of the correlation functions from $pp$ and 
$e^+e^-$ collisions are thus, in my mind, purely accidental and an 
empirical curiosity without physical meaning. {\it The variable 
$Q_{\rm inv}$ should {\em not} be used for fitting HBT data.}]  

Please note also that Eq.~(\ref{corrgauss}) has no factor $\half$ in 
the exponent. If the measured correlator is fitted by a
Gaussian as defined in (\ref{corrgauss}), its $q$-width 
can be directly interpreted in terms of the r.m.s. widths of the 
source in coordinate space. Any remaining factors of $\sqrt{3}$, 
$\sqrt{3}$, or $\sqrt{5}$ (which you can sometimes find in the 
literature) are due to reexpressing the r.m.s. width of the source in 
terms of certain other width parameters chosen for the parametrization 
of the source in coordinate space. The confusion connected with such 
factors is easily avoided by always expressing the source 
parametrization directly in terms of r.m.s. widths.  

Eqs.~(\ref{7}) and (\ref{corrgauss}) would, of course, not be useful 
if the contributions from $\delta S$ and $\delta C$ were not somehow 
small enough to be neglected. This requires a numerical investigation.
It was shown numerically in Ref.~\cite{WSH96} that in typical (and 
even in some not so typical) situations $\delta S$ {\em has a 
negligible influence on the half width of the correlation function}. 
It contributes only weak, essentially unmeasurable structures in 
$C(\bbox{q},\bbox{K})$ at large values of $\bbox{q}$. The reader can 
easily verify this analytically for an emission function with a sharp 
box profile; the results for the exact correlator and the one 
resulting from the Gaussian approximation (\ref{7}) are given in 
\cite{CNH95} and differ by less than 5\% in the half widths; the exact 
correlator has, as a function of $q$, secondary maxima with an 
amplitude below 5\% of the value of the correlator at $q=0$. We have 
checked that similar statements remain even true for a source with a 
doughnut structure, i.e. with a hole in the middle, which was obtained 
by rotating the superposition of two 1-dimensional Gaussians separated 
by twice their r.m.s. widths around their center. The only situation 
where these statements require qualification is if the correlator 
receives contributions from the decay of long-lived resonances; this 
will be discussed in Lecture 3. 

From Eq.~(\ref{corrgauss}) we conclude that the two-particle 
correlation function measures the second order space-time variances
of the emission function. That's it -- finer features of its 
space-time structure (edges, wiggles, holes) cannot be measured with 
two-particle correlations. The variances $\langle \tilde x_\mu \tilde 
x_\nu \rangle$ are in general {\em not} identical with our naive 
intuitive notion of the ``source radius": unless the source is 
stationary and has no $x$-$K$-correlations at all, the variances 
depend on the momentum $\bbox{K}$ of the pair and cannot be 
interpreted in terms of simple overall source geometry. Their correct 
physical interpretation \cite{MS88,AS95,CSH95b} is in terms of 
``lengths of homogeneity" which give, for each pair momentum 
$\bbox{K}$, the size of the region around the point of maximal 
emissivity $\bar x(\bbox{K})$ over which the emission function is 
sufficiently homogeneous to contribute to the correlation function. 
Thus HBT measures ``regions of homogeneity" in the source and their 
variation with the momentum of the particle pairs. As we will see, the 
latter is the key to their physical interpretation.  

\vskip 13pt

\noindent{\bf 3.4.~Gaussian parametrizations for the correlation 
function}

\vskip 13pt

A full characterization of the source in terms of its second order 
space-time variances requires knowledge of the 10 parameters $\langle 
\tilde x_\mu \tilde x_\nu \rangle$. These quantities appear in the 
expression (\ref{corrapp}) for the correlation function but this 
expression still uses all four components of the relative momnetum 
$q^\mu$. However, as already noted only three of the four components 
are independent, due to the mass-shell constraint (\ref{massshell}).
Thus only 6 linear combinations of the variances $\langle 
\tilde{x}_\mu \tilde{x}_\nu \rangle(\bbox{K})$ are actually measurable 
\cite{CNH95}.

If the source is azimuthally symmetric around the beam axis, this 
counting changes as follows: Even if the source is azimuthally 
symmetric in coordinate space, the emission function $S(x,K)$ in phase 
space is for finite $\bbox{K}$ no longer azimuthally symmetric because 
the transverse components $\bbox{K}_\perp$ of the pair momentum 
distinguish a direction transverse to the beam direction. There 
remains, however, a reflection symmetry with respect to the plane 
spanned by $\bbox{K}$ and the beam axis. If we call the direction 
orthogonal to this plane $y$, all mixed variances which are linear in 
$y$ must vanish due to this reflection symmetry, and the correlator 
must be symmetric under $q_y \to -q_y$. Thus only 7 non-vanishing 
variances $\langle \tilde x_\mu \tilde x_\nu \rangle$ survive in 
general, of which, due to the mass-shell constraint (\ref{massshell}) 
only 4 linear combinations are measurable.  

Before the correlator (\ref{corrgauss}) can be fit to experimental 
data, the redundant components of $q$ must first be eliminated from 
the exponent of the Gaussian. At this point it is useful to introduce
a cartesian coordinate system with $z$ along the beam axis and 
$\bbox{K}$ lying in the $x$-$z$-plane. Customarily one labels the 
$z$-component of a 3-vector by $l$ (for {\em longitudinal}), the 
$x$-component by $o$ (for {\em outward}) and the $y$-component by $s$ 
(for {\em sideward}). The above choice of the orientation of the $x$ 
and $y$ axes is natural for azimuthally symmetric collision events or 
event samples because then the transverse components $\bbox{K}_\perp$ 
of the pair momentum distinguish a direction in the transverse plane, 
and it is convenient to orient one of the coordinate axes along this 
direction such that $\bbox{K}$ has only one transverse component: 
 \begin{equation}
 \label{K}
  \bbox{K} = (K_x,K_y,K_z) = (K_\perp,0,K_L)\, .
 \end{equation}
For sources without azimuthal symmetry, e.g. from collisions at 
finite impact parameter which have been selected according to the 
orientation of the collision plane, it is probably more useful to 
orient the $x$ axis along the collision plane; then $\bbox{K}$ will be 
characterized by three parameters, $K_L$, $K_\perp$ and the azimuthal 
angle $\Phi$ of $\bbox{K}_\perp$ relative to the $x$-$z$ collision
plane.

A useful formalism for HBT interferometry of finite impact parameter 
collisions has not yet been developed. I will therefore limit my 
discussion to azimuthally symmetric event samples and exploit the
symmetry to orient the $x$-axis along $\bbox{K}_\perp$. Then from 
(\ref{massshell}) we see that $\beta_s=0$ such that 
 \begin{equation} 
 \label{q0}
   q^0 = \beta_\perp q_o + \beta_l q_l
 \end{equation}
with $\beta_\perp = \vert \bbox{K}_\perp \vert / K^0$ being 
(approximately) the velocity of the particle pair transverse to the 
beam direction while $\beta_l$ is its longitudinal component.

This constraint can now be used in various ways to eliminate the 
redundant $q$-components from the exponent of Eq.~(\ref{corrgauss}).  
But whichever choice one makes, all the $\bbox{K}$-dependent 
parameters (``HBT radii'') in the resulting Gaussian function of $q$ 
can be easily calculated from the variances $\langle \tilde x^\mu 
\tilde x^\nu \rangle$, i.e. by simple quadrature formulae, for 
arbitrary emission functions $S(x,K)$. The relation between the HBT 
parameters and the variances is {\em model-independent}, i.e. it does 
not depend on the form of the emission function $S(x,K)$.  

Here I discuss two specific parametrizations: the standard Cartesian 
one (mostly for historic reasons \cite{pratt,B89,csorgo,CSH95a}), and 
the more physically motivated Yano-Koonin-Podgoretski\u\i\ one 
\cite{CNH95,HTWW96,WHTW96}.  

\vskip 13pt

\noindent{\it 3.4.1.~Standard Cartesian parametrization}

\vskip 13pt

The standard form~\cite{CSH95b,CSH95a} for the parametrization of the 
correlation function is obtained by using (\ref{q0}) to eliminate 
$q^0$ from Eq.~(\ref{corrgauss}). One obtains 
 \begin{equation}
     C(\bbox{q},\bbox{K})
    = 1 + \exp\left[ -\sum_{i,j=s,o,l} R_{ij}^2(\bbox{K})\, q_i\, q_j 
              \right]
 \label{13}
 \end{equation}
where the 6 HBT radius parameters $R_{ij}$ are defined in terms of the 
following variances of the source function~\cite{CSH95b,CSH95a,HB95}: 
 \begin{equation}   
   R_{ij}^2(\bbox{K}) = 
   \langle (\tilde{x}_i-{\beta}_i\tilde{t})
           (\tilde{x}_j-{\beta}_j\tilde{t})\rangle \, ,
   \quad i,j = s,o,l \, .
 \label{14}
 \end{equation}
For an azimuthally symmetric sample of collision events
$C(\bbox{q},\bbox{K})$ is symmetric with respect to $q_s \to -q_s$ 
\cite{CNH95}. Then $R_{os}^2 = R_{sl}^2 = 0$ and 
 \begin{equation}
    C(\bbox{q},\bbox{K})
    = 1 + \exp\left[ - R_s^2(\bbox{K}) q_s^2 - R_o^2(\bbox{K}) q_o^2
                     - R_l^2(\bbox{K}) q_l^2 - 2 R_{ol}^2(\bbox{K}) q_o q_l
              \right] \, ,
 \label{15}
 \end{equation}
with           
 \begin{eqnarray}   
   R_s^2(\bbox{K}) &=& \langle \tilde{y}^2 \rangle \, ,
 \label{16a}\\
   R_o^2(\bbox{K}) &=& 
   \langle (\tilde{x} - \beta_\perp \tilde t)^2 \rangle \, ,
 \label{16b}\\
   R_l^2(\bbox{K}) &=& 
   \langle (\tilde{z} - \beta_l \tilde t)^2 \rangle \, ,
 \label{16c}\\
   R_{ol}^2(\bbox{K}) &=& 
   \langle (\tilde{x} - \beta_\perp \tilde t)
           (\tilde{z} - \beta_l \tilde t) \rangle \, .
 \label{16d} 
 \end{eqnarray}
The cross-term (\ref{16d}) was only recently discovered \cite{CSH95a}.
Clearly these HBT radius parameters mix spatial and temporal 
information on the source in a non-trivial way. Furthermore, since 
they multiply combinations of the components $q^\mu$ which are not 
invariant under longitudinal boosts of the measurement frame, their 
interpretation depends on the frame in which the particle momenta are 
specified. This complicates their physical interpretation. An 
extensive discussion of these parameters, in particular of the meaning 
of the generally non-vanishing cross-term $R_{ol}^2$, 
can be found in Refs.~\cite{CSH95b,CNH95,WSH96,CSH95a,LC96,TH96}, 
where the expressions (\ref{16a})-(\ref{16d}) were analyzed 
analytically and numerically for a large class of (azimuthally 
symmetric) model source functions. Some of these results will be 
discussed in the Lecture 3.

An important observation resulting from these studies is that the 
difference 
 \begin{equation}
 \label{17}
   R_{\rm diff}^2 \equiv  R_o^2 - R_s^2 =
   \beta_\perp^2 \langle \tilde t^2 \rangle - 2 \beta_\perp \langle
   \tilde{x} \tilde t\rangle + (\langle \tilde x^2 \rangle -
   \langle \tilde y^2 \rangle)
 \end{equation}
is generally dominated by the first term on the r.h.s. and thus 
provides access to the lifetime $\Delta t = \sqrt{\langle t^2 \rangle 
- \langle t \rangle^2}$ of the source (more exactly: the duration of 
the particle emission process) \cite{CP91}. However, in relativistic 
heavy-ion collisions, due to rapid expansion of the source one would 
not expect $\langle \tilde t^2 \rangle$ to be generically much larger 
than either $\langle \tilde x^2 \rangle$ or  $\langle \tilde y^2 
\rangle$ unless there is a phase transition to a quark-gluon plasma
and the collision fireball is initiated within a certain 
range of energy densities above the critical energy density where the 
transition occurs \cite{RG96}. In the standard fit one is not 
sensitive to small values of $\Delta t$ since Eq.~(\ref{17}) then 
involves a small difference of two large numbers, each associated with 
standard experimental errors. The factor $\beta_\perp^2 \leq 1$ in 
front of $\langle \tilde t^2 \rangle$ further complicates its 
extraction, in particular at low $K_\perp$ where $\Delta t(\bbox{K})$ 
is usually largest (see below). Successful attempts to determine the
duration of particle emission from HBT measurements have been reported
from low-energy heavy-ion collisions (using 2-proton correlations) 
where the measured lifetimes are very long: $25 \pm 15$ fm/$c$ in 
Ar+Sc collisions at $E/A=80$ MeV \cite{Lisa93} and $1400 \pm 300$ 
fm/$c$ in Xe+Al collisions at $E/A=31$ MeV \cite{Lisa94} (the latter 
is the typical evaporation time of a compound nucleus). Two-pion 
correlations at ultra-relativistic energies ($E/A=200$ GeV) so far 
failed to yield positiveevidence for a non-vanishing emission duration
\cite{NA35,NA44}, except for the heaviest collision system 
Pb+Pb \cite{NA49}, but even there the effective lifetime is only a few 
fm/$c$.

\vskip 13pt

\noindent{\it 3.4.2.~Yano-Koonin-Podgoretski\u\i\ (YKP) parametrization}

\vskip 13pt

The Yano-Koonin-Podgoretski\u\i\ parametrization of the correlation 
function is the generalization to azimuthally symmetric systems 
of the Yano-Koonin parametrization \cite{YK78}. It was first written 
down by M.I.~Podgorestski\u\i\ in 1983 for moving azimuthally 
symmetric, but non-expanding sources \cite{P83}, with $K$-independent 
parameters. Not knowing about this paper, we reinvented it in 
\cite{CNH95}, with $K$-dependent parameters for expanding azimuthally 
symmetric systems. The YKP parametrization is based on an elimination 
in Eq.~(\ref{corrgauss}) of $q_o$ and $q_s$ in terms of $q_{\perp} = 
\sqrt{q_o^2 + q_s^2}$, $q^0$, and $q_3$, using Eq.~(\ref{q0}): 
 \begin{equation}
 \label{18}
   C(\bbox{q},\bbox{K}) =  1 
   +  \exp\left[ - R_\perp^2\, q_{\perp}^2 
                 - R_\parallel^2 \left( q_l^2 - (q^0)^2 \right)
                 - \left( R_0^2 + R_\parallel^2\right) (q\cdot U)^2
                \right]  \, .
 \end{equation}
Like the standard Cartesian form (\ref{15}) it has four $K$-dependent
fit parameters, but now only three of them, $R_\perp(\bbox{K})$, 
$R_\parallel(\bbox{K})$, and $R_0(\bbox{K})$, have dimensions of 
length while the fourth parameter, $U(\bbox{K})$, is a 4-velocity with 
only a longitudinal spatial component: 
 \begin{equation}
 \label{19}
   U(\bbox{K}) = \gamma(\bbox{K}) \left(1, 0, 0, v(\bbox{K}) \right) ,
   \ \ \text{with} \ \
   \gamma = {1\over \sqrt{1 - v^2}}\, .
 \end{equation}
This parametrization has the advantage that the ``YKP radii" 
$R_\perp$, $R_\parallel$, and $R_0$ extracted from such a fit do not 
depend on the longitudinal velocity of the observer system in which 
the correlation function is measured; they are invariant under 
longitudinal boosts. Their physical interpretation is easiest in terms 
of coordinates measured in the frame where $v(\bbox{K})$ vanishes.  
There they are given by \cite{CNH95} 
 \begin{eqnarray}   
   R_\perp^2(\bbox{K}) &=& R_s^2(\bbox{K}) = \langle \tilde{y}^2 \rangle \, ,
 \label{20a} \\
   R_\parallel^2(\bbox{K}) &=& 
   \left\langle \left( \tilde z - \beta_l \tilde x / \beta_\perp
                \right)^2 \right \rangle   
     - \beta_l^2 \langle \tilde y^2 \rangle / \beta_\perp^2
     \approx \langle \tilde z^2 \rangle \, ,
 \label{20b} \\
   R_0^2(\bbox{K}) &=& 
   \left\langle \left( \tilde t -  \tilde x /\beta_\perp
                \right)^2 \right \rangle 
    - \langle \tilde y^2 \rangle / \beta_\perp^2 
    \approx \langle \tilde t^2 \rangle \, ,
 \label{20c}
 \end{eqnarray}
where in the last two expressions the approximation consists of 
dropping terms which were found in \cite{CNH95} to be generically 
small. (A more quantitative discussion of this point will follow in 
Lecture 3.) The first expression (\ref{20a}) remains true in any 
longitudinally boosted frame.

Eq.~(\ref{20c}) shows that the YKP parameter $R_0(\bbox{K})$ 
essentially measures the time duration $\Delta t(\bbox{K}) = 
\sqrt{\langle \tilde t^2 \rangle}$ during which particles of momentum 
$\bbox{K}$ are emitted, in the frame were the YKP velocity 
$v(\bbox{K})=0$. It enters as the leading contribution in $R_0$, is 
fitted directly and no longer obtained as the difference of two large 
fit parameters as in the standard Cartesian fit.

Eqs.~(\ref{20a})-(\ref{20c}) were written down in the special frame 
where $v(\bbox{K})=0$ which we call the {\em Yano-Koonin (YK) frame}.
In Refs.~\cite{CNH95,WHTW96} it is shown that for a large class of 
models this frame essentially coincides with the longitudinal rest 
frame of the fluid cell around the point $\bar x(\bbox{K})$ of maximum 
emissivity at momentum $\bbox{K}$ (i.e. the {\em Longitudinal Saddle 
Point System} LSPS \cite{CL96}). This is true also for sources which 
are not longitudinally boost-invariant and for which the LSPS and the 
LCMS (the {\em Longitudinally CoMoving System} in which the pion pair 
has $\beta_l=0$ \cite{CP91}) do not coincide.  

In general the measurement system will not coincide with the 
($K$-dependent) YK-frame, and the YKP radii will be given by more 
complicated combinations of the space-time variances of the source
expressed in the coordinates of the measurement frame. This is the 
simple result of a Lorentz-boost between the two frames. However, I 
stress that in any frame the YKP parameters can again be easily 
calculated from the second order moments of the source function 
$S(x,K)$, i.e. by simple quadrature. Introducing the notational 
shorthands
 \begin{eqnarray}
 \label{21a}
   A &=& \left\langle \left( \tilde t - \tilde \xi/ \beta_\perp \right)^2 
         \right\rangle \, ,
 \\
 \label{21b}
   B &=& \left\langle \left( \tilde z - \beta_l \tilde \xi / \beta_\perp 
                      \right)^2 
         \right\rangle 
   \, ,
 \\
 \label{21c}
   C &=& \left\langle \left( \tilde t - \tilde \xi/ \beta_\perp \right)
                      \left( \tilde z - \beta_l \tilde \xi / \beta_\perp
                      \right) 
         \right\rangle \, ,
 \end{eqnarray}
where $\tilde \xi \equiv \tilde x + i \tilde y$ and $\langle \tilde 
y\rangle = \langle \tilde x \tilde y \rangle = 0$ for azimuthally 
symmetric sources such that $\langle \tilde \xi^2 \rangle = \langle 
\tilde x^2 - \tilde y^2 \rangle$, one finds in an arbitrary 
longitudinal reference frame 
 \begin{eqnarray}
 \label{22a}
   v &=& {A+B\over 2C} \left( 1 - \sqrt{1 - \left({2C\over A+B}\right)^2}
                       \right) \, ,
 \\
 \label{22b}
   R_\parallel^2 &=& {1\over 2} \left( \sqrt{(A+B)^2 - 4C^2} - A + B
                        \right) = B{-}v C\, ,
 \\
 \label{22c}
   R_0^2 &=& {1\over 2} \left( \sqrt{(A+B)^2 - 4C^2} + A - B 
                        \right) = A{-}v C\, . 
 \end{eqnarray}
The Yano-Koonin velocity $v$ is zero in the frame where the 
expression (\ref{21c}) for $C$ vanishes \cite{CNH95}; this fixes also 
the sign in front of the square root in (\ref{22a}). For small values 
of $C$ the Yano-Koonin velocity is given approximately by 
 \begin{equation}
 \label{23}
   v \approx {C\over A+B} 
     \approx {\langle \tilde z \tilde t \rangle \over
              \langle \tilde t^2 \rangle + \langle \tilde z^2 \rangle} 
 \, , 
 \end{equation}
where in the second approximation we again neglected generically small 
terms \cite{CNH95} proportional to $\langle \tilde z \tilde x\rangle$, 
$\langle \tilde x \tilde t \rangle$, and $\langle \tilde x^2 - \tilde 
y^2 \rangle$. The accuracy of the approximate expression (\ref{23}) 
for $v(\bbox{K})$ was tested numerically in \cite{WHTW96} and found to 
be excellent in the situations discussed below. In the same limit the 
expressions for $R_0^2$ and $R_\parallel^2$ simplify to $R_0^2 \approx 
A$ and $R_\parallel^2 \approx B$, in agreement with (\ref{20b}) and 
(\ref{20c}).  

Since the standard Cartesian and YKP parametrizations are 
mathematically equivalent (being simply based on a different choice of 
independent $q$-components), the resulting HBT parameters obey simple 
relations \cite{HTWW96}: 
 \begin{eqnarray}
 \label{24}
   R_s^2 &=& R_\perp^2\, ,
 \\
 \label{24a}
   R_{\rm diff}^2 &=& R_o^2 - R_s^2 = \beta_\perp^2 \gamma^2 
             \left( R_0^2 + v^2 R_\parallel^2 \right) \, ,
 \\
 \label{24b}
   R_l^2 &=& \left( 1 - \beta_l^2 \right) R_\parallel^2 
             + \gamma^2 \left( \beta_l-v \right)^2
             \left( R_0^2 + R_\parallel^2 \right) \, ,
 \\
 \label{24c}
   R_{ol}^2 &=& \beta_\perp \left( -\beta_l R_\parallel^2 
             + \gamma^2 \left( \beta_l-v \right)^2
             \left( R_0^2 + R_\parallel^2 \right) \right) \, .
 \end{eqnarray}
Although a mathematical triviality, these relations provide a powerful 
consistency check on the experimental fitting procedure, of similar 
value as the relation \cite{CNH95,WSH96} $\lim_{K_\perp \to 0} 
(R_o(\bbox{K}) - R_s(\bbox{K})) = 0$ which results from azimuthal 
symmetry.  

%
%
%

\vskip 26pt

\noindent{\bf 4.~LECTURE 3: HBT CORRELATIONS FOR EXPANDING SOURCES}

\vskip 13pt

In this third lecture I will present a quantitative discussion of the 
HBT parameters, both in the standard Cartesian and in the YKP fits.  
The emphasis will be on their $K$-dependence because, as discussed in 
the previous two lectures, only a careful study of this pair-momentum 
dependence permits a separation of the geometric from the dynamic 
aspects of the source. You probably remember that at the beginning of 
Lecture 2 I stressed that a model-independent reconstruction of the 
emission function from HBT measurements is not possible. Therefore 
any quantitative discussion must necessarily occur within the 
framework of specific source models. 

I will choose a relatively simple analytical parametrization of the 
emission function which was first suggested by T. Cs\"org\H o and B. 
L\"orstad in 1994 in an unpublished preprint (see also 
\cite{CL96,CSH95a,CSH95b,CNH95}) and which incorporates many of the 
(as we believe) relevant physical features of the typical sources 
created in relativistic nuclear collisions: approximate thermalization 
at decoupling, finite transverse and longitudinal extension of the 
source, collective expansion in both the longitudinal and transverse 
directions, and a finite duration of the particle emission process.  
All these features are controlled by parameters which can be freely 
tuned, thus allowing for extensive parameter studies 
\cite{CNH95,WSH96,WHTW96,TH96} which have given us a good intuitive 
understanding of which properties of the source are important for the 
correlation function and which aren't, and how to look in the 
correlation function for specific space-time properties of the 
emitter.  

\vskip 13pt

\noindent{\bf 4.1.~A simple model for an expanding source}

\vskip 13pt

Let us consider the following model for the emission function 
\cite{CNH95,CL96}: 
 \begin{equation}
   S(x,K) = {2J+1\over (2\pi)^3}\, 
            K\cdot n(x)\, 
            \exp \left[- {K \cdot u(x) - \mu (x)\over T(x)}\right] \,
            H(x)
 \label{2.1}
 \end{equation}
with
 \begin{equation}
   H(x) = {1\over \pi \Delta \tau}\,
          \exp \left[- {r^2 \over 2 R^2}
                     - {(\eta- \eta_0)^2 \over 2 (\Delta \eta)^2}
                     - {(\tau-\tau_0)^2 \over 2(\Delta \tau)^2}
               \right] 
 \label{2.1a}
 \end{equation}
and
 \begin{equation}
   K\cdot n(x) = M_\perp \cosh(\eta-Y) \, .
 \label{2.1b}
 \end{equation}
The factor $2J+1$ counts the spin degeneracy of the observed particle 
species and is included because the detectors usually do not identify 
the polarization of the measured particles. There is no such factor 
for isospin because of the requirement that the two particles in the pair 
be identical, i.e. have the, for example, the same electric charge. 
The Lorentz covariant Boltzmann factor $\exp[-( K{\cdot}u(x) - 
\mu(x))/T(x)]$ implements the assumption of local thermal equilibrium 
of the emitted particles at freeze-out, with local temperature $T(x)$ 
and chemical potential $\mu(x)$, and the collective expansion of the 
source with hydrodynamic flow 4-velocity $u_\mu(x)$. I will here take
$T$ and $\mu$ as constants and refer to Refs.~\cite{CL96,TH96} for an 
investigation of the effects of temperature gradients. The factor 
$H(x)$ describes the geometric properties of the source; it can be 
interpreted as a space-time modulation of the fugacity 
$\exp[\mu(x)/T(x)]$. Space-time is parametrized by {\em longitudinal 
proper time} $\tau= \sqrt{t^2-z^2}$ and {\em space-time rapidity} 
$\eta = {1 \over 2} \ln[(t+z)/(t-z)]$ in the longitudinal and temporal 
directions, and by $r= \sqrt{x^2+y^2}$ and $\phi$ in the transverse 
directions. The volume element is then $d^4x = \tau\, d\tau\, d\eta\, 
r\, dr\, d\phi$. The Gaussian factors $\exp[-r^2/(2R^2)]$ and $\exp[-
(\eta-\eta_0)^2/(2(\Delta\eta)^2)]$ provide smooth geometric limits 
for the source in the transverse and longitudinal directions, scaled 
by $R$ and $L=\tau\Delta\eta$, respectively. The function $H(x)$ is 
normalized to the total comoving 3-volume according to 
 \begin{equation}
   \int d^4x\, H(x) = \pi r_{\rm rms}^2 \cdot 2\tau_0 \eta_{\rm rms}
 \label{normH}
 \end{equation}
where
 \begin{equation}
  r^2_{\rm rms} = 2R^2 = x^2_{\rm rms} + y^2_{\rm rms}\, ,
  \qquad
  \eta_{\rm rms} = \Delta\eta
 \label{normH1}
 \end{equation}
are the r.m.s. expectation values for the transverse radius $r$ and 
for $\eta$, respectively. (The r.m.s. widths for $x$ and $y$ are each 
given by $R$.) If the Gaussians in $H(x)$ were replaced by box 
functions, the equivalent box dimensions (with the same r.m.s. radii) 
would be $\tilde R=2R$, $\tilde\eta = \sqrt{3} \Delta\eta$. (Here you 
see the ``unneccesary" factors $\sqrt 2$ and $\sqrt 3$ mentioned 
before!) 

$K{\cdot}n(x)$ is the flux factor through the freeze-out hypersurfaces 
whose normal direction is given by the unit vector $n_\mu(x)$. In our 
model these hypersurfaces are for simplicity assumed to be surfaces of 
constant longitudinal proper time $\tau$, parametrized by surface 
coordinates $\Sigma_{(\tau)}(x) = (\tau \cosh\eta, r \cos\phi, r 
\sin\phi, \tau \sinh\eta)$. The last factor in $H(x)$ provides a 
Gaussian smearing of the proper time around the mean values $\tau_0$ 
with dispersion $\Delta\tau$, thereby implementing particle emission 
over a finite time interval of order $\Delta\tau$ in the local 
comoving frame. With these assumptions the flux factor reduces 
\cite{CNH95} to the form given in Eq.~(\ref{2.1b}). For $\Delta \tau 
\to 0$ the Gaussian in $\tau$ approaches a $\delta$-function centered 
at $\tau_0$, simulating instantaneous freeze-out at constant proper 
time as often implemented in hydrodynamical situations. 

[Let me make some side remarks on hydrodynamical simulations here,
because they are a very popular tool for computing the space-time 
dynamics of the reaction zone in heavy-ion collisions. In 
hydrodynamics freeze-out is always assumed to occur on a sharp 
hypersurface (not a smeared one as here), and one writes the emission 
function in the form \cite{marb} 
 \begin{equation}
   S(x,K) = \frac{2J+1}{(2\pi)^3}
   \int_\Sigma 
   \frac{K^\mu d^3\!\sigma_\mu(x')\,\delta^{(4)}(x-x')}
   {\exp[(K{\cdot}u(x')-\mu(x'))/T(x')] \pm 1 } \;,
 \label{hydro}
 \end{equation}
where $d^3\!\sigma_\mu(x')$ is the normal vector on the freeze-out
surface $\Sigma(x')$, and we have correctly accounted for the quantum 
statistical effects in the local thermal distribution functions. The
latter are unimportant for heavy particles but should, in a 
quantitative comparison with data, be included for pions -- at least 
in the single-particle spectrum. Inserting the ansatz (\ref{hydro})  
into the expression (\ref{spectrum}) for the single-particle spectrum
one obtains the well-known Cooper-Frye formula \cite{coop} 
 \begin{equation}
   E_K {dN\over d^3K} = 
   \int_\Sigma K^\mu d^3\!\sigma_\mu(x)\, f(x,p)
 \label{coop}
 \end{equation}
with the local equilibrium distribution function
 \begin{equation}
   f(x,p) = \frac{2J+1}{(2\pi)^3}\frac{1}
   {\exp[(K{\cdot}u(x)-\mu(x))/T(x)] \pm 1}\;.
 \label{distr}
 \end{equation}
For the numerator of the correlator in (\ref{correlator}) one obtains 
 \begin{equation}
   \int_\Sigma K^\mu d^3\!\sigma_\mu(x)\,  K^\nu d^3\!\sigma_\nu(y)\,
     f(x,K)\,f(y,K)\,\exp[iq{\cdot}(x-y)] \, ,
 \label{numerator}
 \end{equation}
similar, but not identical with to the one given in
\cite{sinu}. In \cite{sinu} each of the two distribution 
functions under the integral featured on-shell arguments $p_a$ and
$p_b$, respectively, instead of the common (off-shell) average
argument $K$ as in (\ref{numerator}). This error in \cite{sinu} can be
traced back to an inaccurate transition from finite discrete volumes
along the freeze-out surface $\Sigma$ to the continuum limit 
\cite{chap}. Taking over this inaccuracy produces (in particular for 
very rapidly expanding sources) unphysical oscillations of the 
correlation function around unity at large values of $\bbox{q}$ (see 
e.g. \cite{sriv}) which are inconsistent with the manifestly positive 
definite nature of the exchange term noted in Lecture 1. Recently 
Aichelin \cite{aich2} pointed out that, for technical reasons, the 
same erroneous assumption is made in all simulations of HBT 
correlations using Monte Carlo event enerators. This problem should 
certainly receive more attention in the future.] 

Since according to Eq.~(\ref{Konshell}) the time-component of the pair 
momentum may be approximated by the on-shell value $K_0 = E_K = 
\sqrt{m^2 + {\bf K}^2}$, the pair momentum $K$ can be parametrised 
in terms of the momentum rapidity $Y = {1\over 2} \ln[(1+\beta_l)/(1-
\beta_l)]$ and the transverse mass $M_\perp = \sqrt{m^2 + K_\perp^2}$, 
  \begin{equation}
    \label{2.2}
    K^\mu = (M_\perp\cosh{Y}, K_\perp, 0, M_\perp\sinh{Y}).
  \end{equation}
We implement longitudinal and azimuthally symmetric transverse expansion 
of the source by parametrising the flow velocity in the form
 \begin{equation}
 \label{2.3}
   u^{\mu}(x) = \left( \cosh \eta_l(\tau,\eta) \cosh \eta_t(r), \,
                     {x\over r}\sinh \eta_t(r),  \,
                     {y\over r}\sinh \eta_t(r),  \,
                     \sinh \eta_l(\tau,\eta) \cosh \eta_t(r) \right) .
 \end{equation}
For the longitudinal flow rapidity we take $\eta_l(\tau,\eta) = \eta$ 
independent of $\tau$, i.e. we assume a Bjorken scaling profile \cite{BJ83}
$v_l=z/t$ in the longitudinal direction. The growth of the 
longitudinal flow velocity in the longitudinal direction is then 
automatically limited by the Gaussian in $\eta$ in (\ref{2.1a}). 
For the transverse flow rapidity we take a linear profile of strength 
$\eta_f$:  
 \begin{equation}
 \label{2.4}
  \eta_t(r) = \eta_f \left( {r \over R} \right)\, .
 \end{equation}
The scalar product in the exponent of the Boltzmann factor can then be 
written as
  \begin{equation}
    \label{2.5}
    K\cdot u(x) = M_\perp \cosh(\eta - Y) \cosh\eta_t(r) 
                - K_\perp {x\over r} \sinh\eta_t(r) \, ,
  \end{equation}
Please note that for non-zero transverse momentum $K_\perp$, a finite 
transverse flow breaks the azimuthal symmetry of the emission function 
via the second term in (\ref{2.5}). For $\eta_f=0$ the source has no 
explicit $K_\perp$-dependence, and $M_\perp$ is the only relevant scale. 
As will be discussed later this gives rise to perfect 
$M_\perp$-scaling of the YKP radius parameters in the absence of 
transverse flow, which is again broken for non-zero transverse flow 
\cite{HTWW96a}.  

Besides $\eta_f$, the model parameters are the freeze-out temperature 
$T$, the transverse geometric (Gaussian) radius $R$, the average 
freeze-out proper time $\tau_0$ as well as the mean proper emission 
duration $\Delta \tau$, the centre of the source rapidity 
distribution $\eta_0$, and the (Gaussian) width of the space-time 
rapidity profile $\Delta \eta$. A rough spatial picture of the source 
at various fixed coordinate times can be gleaned from Figs.~1 and 2 in 
Ref.~\cite{CN96} to which I would like to refer those readers having 
trouble with visualizing Gaussians in $\eta$ and $\tau$ in regular
Cartesian coordinates. Although the source in Ref.~\cite{CN96} has 
sharp edges in space and time whereas ours is smoothed by Gaussian 
profiles, the essential space-time features of the sources are very 
similar.  

The calculations presented below were done for pions 
($m = m_{\pi^\pm}=139$ MeV/$c^2$) and kaons ($m=m_{K^\pm}=494$ 
MeV/$c^2$) and (except were noted otherwise) for the fixed set of 
parameters $T=140$ MeV, $R=3$ fm, $\tau_0=3$ fm/$c$, $\Delta\tau=1$ 
fm/$c$, $\Delta\eta=1.2$, and $\eta_0=0$ (i.e. all our rapidities will 
be given relative to the rapidity $\eta_0$ of the c.m. of the source).
The strength $\eta_f$ of the transverse flow will be varied 
systematically to investigate its effects on the HBT parameters. A 
detailed discussion of how the variation of some of the other 
parameters affects the correlation function can be found in 
Refs.~\cite{WHTW96,TH96}.

\vskip 13pt

\noindent{\bf 4.2.~$K$-dependence of the HBT radii}

\vskip 13pt

\noindent{\it 4.2.1.~Analytical approximations -- HBT radii as 
lengths of homogeneity}

\vskip 13pt

For a qualitative understanding of the physical origin of the
pair-momentum dependence of the HBT parameters it is instructive to 
start from their model-independent expressions in terms of space-time
variances and evaluate the latter approximately analytically.
For the standard Cartesian fit parameters (\ref{16a})-(\ref{16d}) 
this was done, at different levels of accuracy, in 
Refs.~\cite{CSH95b,WSH96,AS95,CL96} by exploiting the method of 
saddle-point integration for the evaluation of the variances. This 
method was introduced by Makhlin and Sinyukov \cite{MS88} in the 
context of infinitely long sources with boost-invariant longitudinal 
expansion where they used it to show that the longitudinal HBT radius 
$R_l$ is
\break

 \vspace*{15.5cm}
 \includegraphics{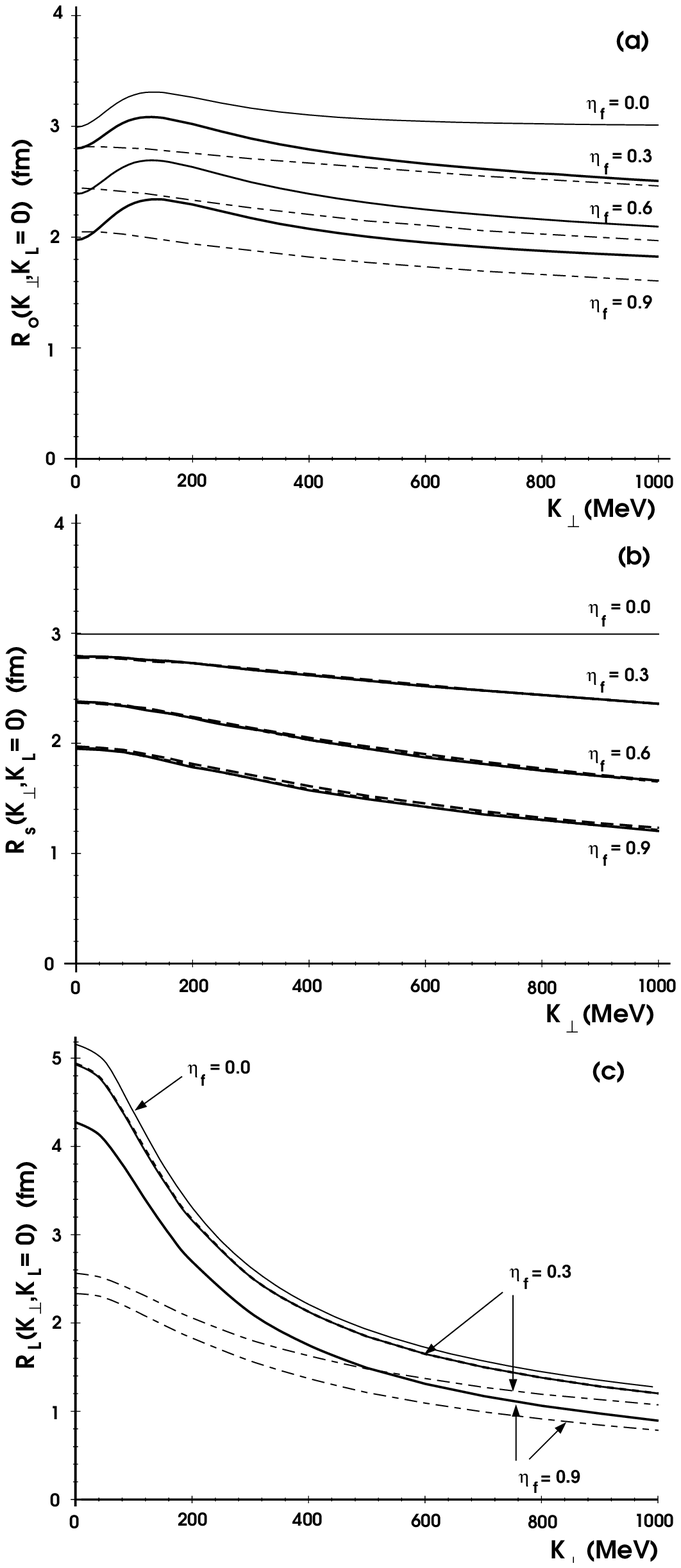}
 \begin{center}
 \begin{minipage}[t]{13cm}
 \noindent {\footnotesize \bf Fig.1.  \rm
$K_\perp$-dependence of the standard Cartesian HBT radii $R_o$ (top), 
$R_s$ (middle), and $R_l$, evaluated in the LCMS (the frame where 
$K_L=0$), for an infinitely long source ($\Delta\eta \to \infty$ in 
Eqs.~(\protect\ref{2.1},\protect\ref{2.1a})) with boost-invariant 
longitudinal expansion and a linear transverse flow rapidity profile.  
The strength of the tranverse flow $\eta_f$ is varied between 
$\eta_f=0$ and $\eta_f=0.9$ as indicated. The duration 
parameter $\Delta\tau$ was set to 0 (locally instantaneous 
freeze-out).  For such a source the cross-term $R_{ol}$ vanishes 
identically in the LCMS.  Solid lines represent results from a full 
numerical evaluation of the integrals in 
Eqs.~(\protect\ref{16a}-\protect\ref{16d}). Dashed lines give the 
results from the leading-order saddle-point approximation to these 
integrals. One sees that the saddle-point integration misses the rise 
of $R_o$ at small $K_\perp$ (i.e. the contribution from the finite 
duration of particle emission in the LCMS frame, see text) and gives a 
very bad approximation to $R_l$ at small $K_\perp$. The solid line
for $R_l$ for $\eta_f=0$ is given analytically by
$R_l = \tau_0 \sqrt{T/M_\perp} \sqrt{K_2(M_\perp/T)/K_1(M_\perp/T)}$
\protect\cite{HB95} while the saddle-point approximation (dashed) 
gives for $\eta_f=0$ the Makhlin-Sinyukov formula $R_l = \tau_0 
\sqrt{T/M_\perp}$ \protect\cite{MS88}.  
(Figure taken from Ref.~\protect\cite{WSH96}.)} 
 \end{minipage} 
 \end{center} 

\noindent
finite and determined by the inverse of the longitudinal 
velocity gradient, noting for the first time that $R_l$ has the 
property of a ``longitudinal length of homogeneity" in the source 
rather than being related to the longitudinal geometric size of the 
entire source. It was later observed \cite{CSH95b,WSH96} that not all 
of the important qualitative features of the $K$-dependence of the HBT 
parameters can be obtained from the leading term in the saddle point 
approximation (see Fig.~1). In particular, in the presence of 
transverse flow the saddle point moves away from the beam axis $r=0$, 
and this must be taken into account in order to obtain reasonable 
approximations \cite{WSH96}. Unfortunately, this renders the whole 
procedure rather cumbersome \cite{WSH96}, and in the end, e.g. for a 
quantitative comparison with data, a full numerical evaluation of the 
integrals for the variances cannot be avoided.  

In spite of their unreliability on a quantitative level, the analytical 
results from saddle-point integration are still very instructive on a 
qualitative level. I will here discuss the leading results for a 
longitudinally finite ($L=\tau_0\Delta\eta$) source of type 
(\ref{2.1}) in the limit $\Delta\tau=0$, in the LCMS (i.e. in the 
Longitudinally CoMoving System where the pairs have $K_L=0$). 
One finds \cite{CSH95b}
 \begin{eqnarray}
 \label{Rs}
    R_s^2(M_\perp,K_L=0) &=& R_*^2 \, ,
 \\
 \label{Ro}
    R_o^2(M_\perp,K_L=0) &=& R_*^2
                          + \beta_\perp^2 (\Delta t_*)^2\, ,
 \\
 \label{Rl}
    R_l^2(M_\perp,K_L=0) &=& L_*^2 \, ,
 \\
 \label{Rol}
    R_{ol}^2(M_\perp,K_L=0) &=& {\beta_\perp\, Y \over \sqrt{2}}\, 
    L_*\, \Delta t_* \left(L_*^2\over L^2\right) \, ,
 \end{eqnarray}
where $R_*$, $L_*$, and $\Delta t_*$ are functions of $M_\perp$ 
defined by
 \begin{eqnarray}
 \label{Rstar}
   {1\over R_*^2} = {1\over R^2} + {1\over R_H^2}\, , 
 \\
 \label{Lstar}
   {1\over L_*^2} = {1\over L^2} + {1\over L_H^2}\, , 
 \\
 \label{tstar}
   \Delta t_* = \sqrt{2} 
   \left( \sqrt{ \tau_0^2 + L_*^2} - \tau_0 \right) \, , 
 \end{eqnarray}
with the transverse and longitudinal ``dynamical lengths of 
homogeneity" 
 \begin{eqnarray}
 \label{RH}
   R_H(M_\perp) = {R\over \eta_f}\, \sqrt{{T\over M_\perp}}
   = {1\over \partial \eta_t(r)/\partial r} \, \sqrt{{T\over M_\perp}}\, ,
 \\
 \label{LH}
   L_H(M_\perp) = \tau_0\, \sqrt{{T\over M_\perp}}
   = {1\over \partial{\cdot}u_l} \, \sqrt{{T\over M_\perp}}\, .
 \end{eqnarray}
Strictly speaking, expression (\ref{RH}) is only valid for weak
transverse flow $\eta{\rm f}\ll 1$. In (\ref{LH}) we have defined
the longitudinal flow 4-velocity $u_l = (\cosh\eta_l,0,0,\sinh\eta_l)
= (\cosh\eta,0,0,\sinh\eta)$.

The physical interpretation of these results is quite interesting: 
in addition to geometry (implemented by the Gaussian cutoff factors in 
the function $H(x)$), dynamics affects the HBT radii through the 
dynamical homogeneity lenghts $R_H$, $L_H$. The latter are inversely 
proportional to the gradients of the expansion velocity field in the 
respective direction, but smeared by a thermal smearing factor 
$\sqrt{T/M_\perp}$ resulting from the random thermal motion of the 
particles around the fluid velocity. The HBT radii are determined by 
the shorter of the two (geometric or dynamic) lengths scales. In the 
absence of random thermal motion ($T\to 0$) any velocity gradient in 
the system would lead to a vanishing dynamical length of homogeneity 
and consequently to vanishing HBT radii. At finite $T$, the dynamical 
smearing decreases with increasing transverse mass $M_\perp$, leading 
to a decrease of the HBT radii at large $M_\perp$. (It turns out that 
the $\sqrt{1/M_\perp}$-scaling of the HBT radii at large $M_\perp$ 
suggested by these analytical expressions is unreliable and a 
consequence of the saddle-point approximation \cite{WSH96}. A 
numerical analysis \cite{HTWW96a,WHTW96} shows that the power of 
$M_\perp$ itself, by which the HBT radii decrease for increasing 
$M_\perp$, is proportional to the expansion velocity gradient.) 

The following intuitive picture results from these considerations:
if the expansion velocity is small, i.e. all velocity gradients can be 
essentially neglected over the range where the geometric Gaussians in 
$H(x)$ are large, then HBT measures the geometric parameters $R$, $L$
which tell you where the function $H(x)$ (and thus the whole emission 
function $S(x,K)$) gets cut off. If, on the other hand, the velocity 
gradients are large, they effectively cut off the emission function at 
a distance $R_H$ resp. $L_H$ from the saddle point, and the matter 
outside these homogeneity regions decouples from the correlator 
because it cannot contribute particles with sufficiently small 
relative momenta to see the effects of quantum statistics. This 
explains my statement above that HBT does in general not measure the 
geometry of the source, but rather the regions of homogeneity inside 
the source at a given wavelength $1/K$.

Of course, a space-time dependence of the temperature field $T(x)$ 
can induce additional gradients into the emission function and thus 
affect the size of the regions of homogeneity in the source and the 
HBT radii. This was investigated in some detail in 
Refs.~\cite{CL96,TH96} to which I refer the interested reader. 

The last point to be discussed is the origin of the quantity $\Delta 
t_*$ in Eq.~(\ref{Ro}) for $R_o^2$. Comparing with Eq.~(\ref{17}) we 
see that it has the meaning of an effective source lifetime. But where 
does it come from, since we set the width $\Delta\tau$ of the proper 
time distribution to zero? This apparent paradox has an interesting 
answer which is reflected in the mathematical structure of 
Eq.~(\ref{tstar}): since the correlator receives non-vanishing 
contributions from a longitudinal region of homogeneity of size $R_l = 
L_*$, it probes emission from different points $z$ at different times 
$t = \sqrt{\tau_0^2 + z^2}$ along the proper-time hyperbola $\tau = 
\tau_0$, with maximal range $-L_* \lapp z \lapp L_*$. Thus even for 
sharp freeze-out at constant proper time the correlator sees a 
non-vanishing effective lifetime in the fixed observer frame (here the 
LCMS) which is in principle measurable via the difference $R_o^2-
R_s^2$. Since $L_*$ is a decreasing function of $M_\perp$, so is 
$\Delta t_*$, and for large $M_\perp$ the difference $R_o^2-R_s^2$ 
vanishes. (If $\Delta\tau$ were nonzero, the difference would at large 
$M_\perp$ approach the limit $(\Delta\tau)^2$.) 

\vskip 13pt

\noindent{\it 4.2.2.~Cartesian HBT radii in the CMS and LCMS}

\vskip 13pt

In Fig.~2 I show the HBT radius parameters from the standard Cartesian fit
(\ref{15}) for pion pairs with c.m. rapidity $Y=1.5$ where the fit of the
correlator is done in the CMS \cite{TH96}. The different thick curves 
correspond to different strengths $\eta_f$ of the transverse flow. Without 
transverse flow $R_s$ is $M_\perp$-independent because the we consider 
(\ref{2.1}) for constant temperature and neglect possible transverse 
temperature gradients. As the transverse flow increases, $R_s$ develops an 
increasing dependence on $M_\perp$. As shown in Fig.~6 below it can 
be approximated by an inverse power law, with the power increasing 
monotonously with $\eta_f$ \cite{WSH96,WHTW96}.  

$R_l$ features a very strong $M_\perp$-dependence even without 
transverse flow, due to the strong longitudinal expansion of the 
source. It can also be described by an inverse power law, with a 
larger power $\simeq 0.55$, in rough agreement with the approximate 
$\sqrt{T/M_\perp}$-scaling law suggested in \cite{MS88} (see, however, 
\cite{WSH96,HB95} for a more quantitative discussion). The increase of 
$R_o$ at small $M_\perp$ is due to the contribution (\ref{17}) from 
the effective lifetime. As seen in Fig.~5 below, in the YK frame 
(source rest frame) the latter is of order 2.5 fm/$c$ at small 
$M_\perp$; Fig.~2b shows that its effect on $R_o$ compared to $R_s$ in 
the CMS is much smaller (and thus more difficult to measure). Fig.~2d 
shows that the cross-term is small in the CMS but non-zero. It 
vanishes at $K_\perp=0$ by symmetry and also becomes small again at 
large $K_\perp$.  

\vspace*{9cm}
\includegraphics{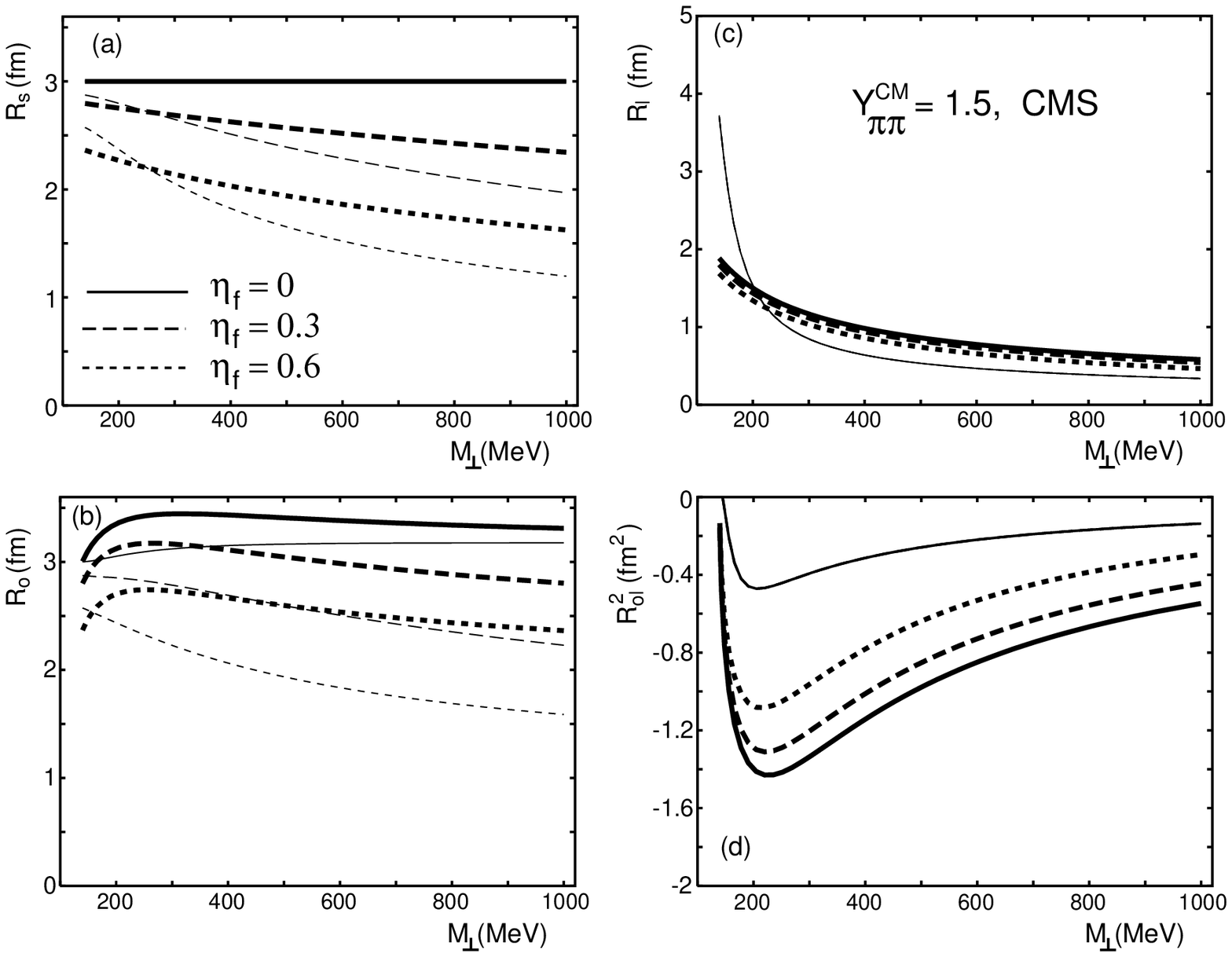}
\begin{center}
\begin{minipage}[t]{13cm}
\noindent {\footnotesize \bf Fig.2.  \rm
The standard Cartesian parameters $R_s$ (a), $R_o$ (b), $R_l$ (c),
and $R_{ol}^2$ (d) in the CMS for pion pairs with c.m. rapidity $Y=1.5$,
as functions of $M_\perp$ for 3 different values for the transverse 
flow $\eta_f$. The thick lines are exact numerical results from 
Eqs.~(\protect\ref{16a}-\protect\ref{16d}), the thin lines are obtained 
from the analytical approximations given in Ref.~\protect\cite{CL96}. 
(Figure taken from Ref.~\protect\cite{TH96}.)}
\end{minipage}
\end{center}

The thin lines in Fig.~2 show for comparison approximate results for 
the HBT radii calculated from the approximate analytical results given 
in Ref.~\cite{CL96} which were derived by evaluating 
Eqs.~(\ref{16a}-\ref{16d}) by saddle point integration. It is clear 
that this method fails here (see Ref.~\cite{WSH96} for a quantitative 
discussion of this approximation), and that the analytical expressions 
should not be used for a quantitative analysis of HBT data.  

Fig.~3 shows the same situation as Fig.~2, but now all HBT radii are 
evaluated in the LCMS (longitudinally comoving system \cite{CP91}) which 
moves with the pair rapidity $Y=1.5$ relative to the CMS. A comparison with 
Fig.~2 shows the strong reference frame dependence of the standard HBT 
radii. In particular, the cross-term changes sign and is now much larger.
The analytical approximations from Ref.~\cite{CL96} work much better in 
the LCMS \cite{CL96}, but for $R_o$ and $R_{ol}^2$ they are still not 
accurate enough (in particular in view of the delicate nature of the 
lifetime effects on $R_o$).

\vskip 13pt

\noindent{\it 4.2.3.~The Yano-Koonin velocity}

\vskip 13pt

Fig.~4 shows (for pion pairs) the dependence of the YK velocity on the 
pair momentum ${\bf K}$. In Fig.~4a we show the YK rapidity $Y_{_{\rm 
YK}} = \frac 12 \ln[(1+v)/(1-v)]$ as a function of the pair rapidity 
$Y$ (both relative to the CMS) for different values of $K_\perp$, in 
Fig.~4b the same quantity as a function of $K_\perp$ for different 
$Y$.  Solid lines are without transverse flow, dashed lines are for 
$\eta_f=0.6$.  For large $K_\perp$ pairs, the YK rest frame approaches 
the LCMS (which moves with the pair rapidity $Y$); in this limit all 
pairs are thus emitted from a small region in the source which moves 
with the same longitudinal velocity as the pair. For small $K_\perp$ 
the YK frame is considerably slower than the LCMS; this is due to the 
thermal smearing of the particle velocities in our source 
\break

\vspace*{9cm}
\includegraphics{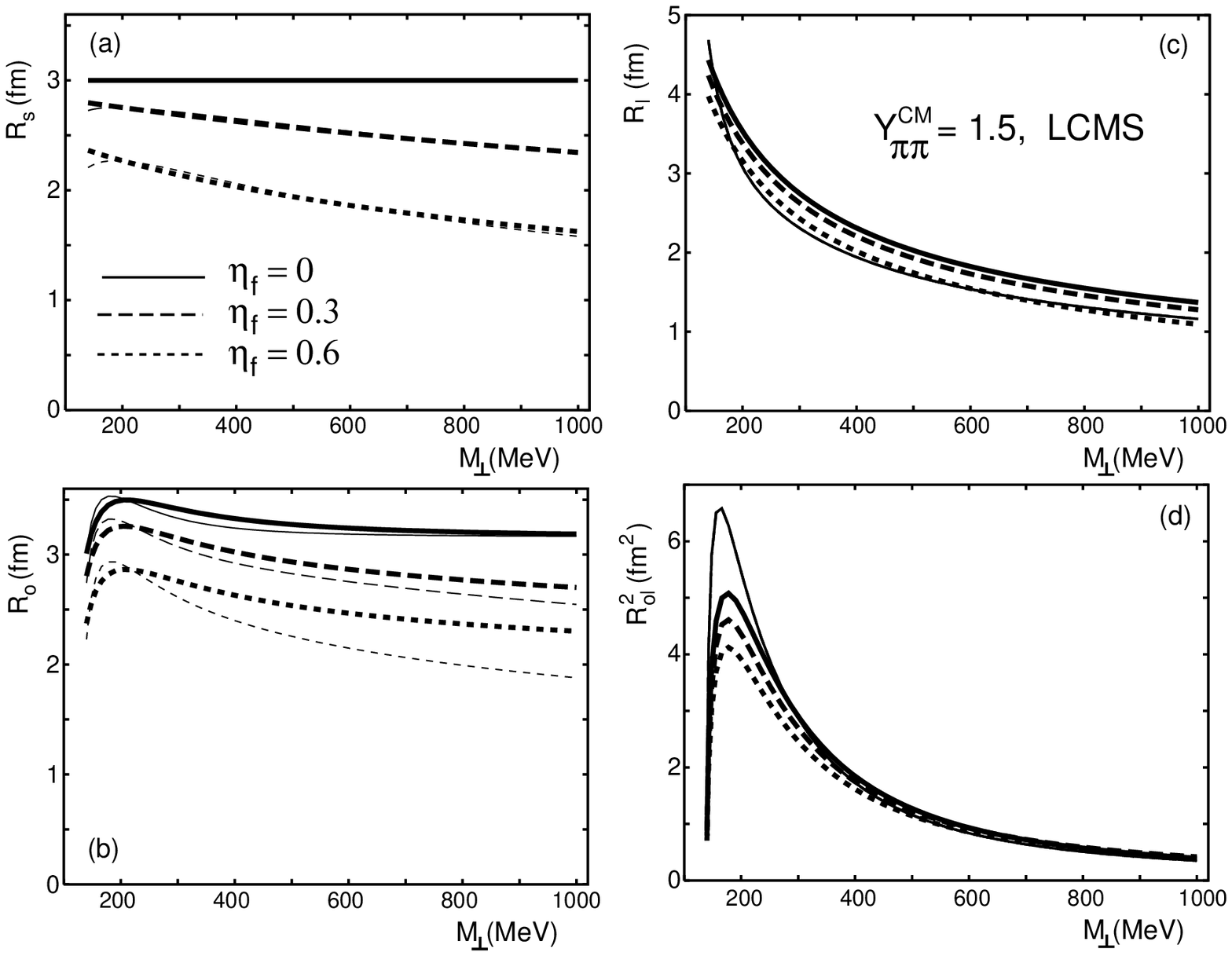}
\begin{center}
\begin{minipage}[t]{13cm}
\noindent {\footnotesize \bf Fig.3.  \rm
Same as Fig.~2, but now evaluated in the LCMS. Please note the change of sign 
and magnitude of the cross-term. 
(Figure taken from Ref.~\protect\cite{TH96}.)}
\end{minipage}
\end{center}

\noindent
around the 
local fluid velocity $u^\mu(x)$ \cite{WHTW96}. The linear relationship 
between the rapidity $Y_{_{\rm YK}}$ of the Yano-Koonin frame and the 
pion pair rapidity $Y$ is a direct reflection of the boost-invariant 
longitudinal expansion flow \cite{HTWW96}. For a non-expanding source 
$Y_{_{\rm YK}}$ would be independent of $Y$. Additional transverse 
flow is seen to have nearly no effect. The dependence of the YK 
velocity on the pair rapidity thus measures directly the longitudinal 
expansion of the source and cleanly separates it from its transverse 
dynamics. A detailed discussion of these features is given in 
Ref.~\cite{WHTW96} where it is also shown that the YK velocity is 
always very close to the velocity of the Longitudinal Saddle Point 
System LSPS (i.e. to the longitudinal velocity of the fluid element 
around the point of maximal emissivity at momentum $K$). This last 
observation establishes the usefulness of the YK velocity (which can 
be directly extracted from an YKP fit to the data) as a measure for 
the longitudinal expansion velocity of the source. 

\vspace*{6cm}
\includegraphics{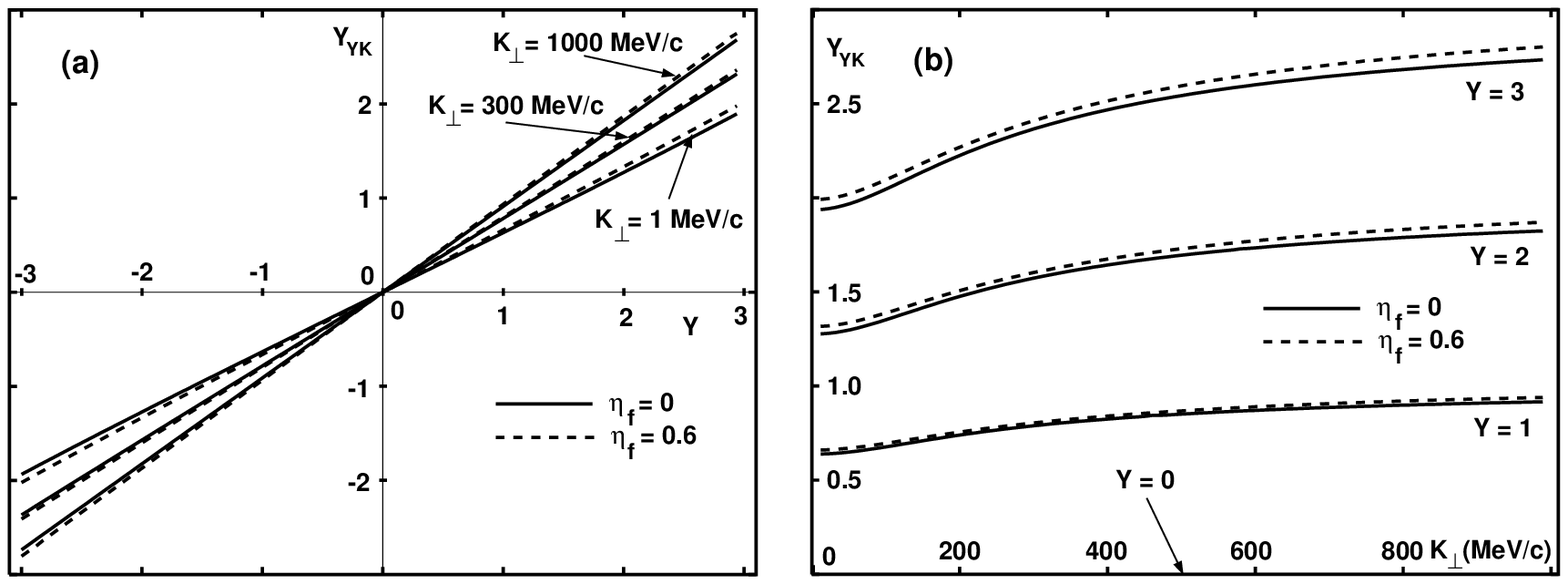}
\begin{center}
\begin{minipage}[t]{13cm}
\noindent {\footnotesize \bf Fig.4.  \rm
(a) The Yano-Koonin rapidity for pion pairs, as a function of the pair 
c.m. rapidity $Y$, for various values of $K_\perp$ and two values for 
the transverse flow $\eta_f$. (b) The same, but plotted against $K_\perp$ 
for various values of $Y$ and $\eta_f$.
(Figure taken from Ref.~\protect\cite{HTWW96}.)}
\end{minipage}
\end{center}

\vskip 13pt

\noindent{\it 4.2.4.~YKP radii: $M_\perp$-scaling and transverse flow}

\vskip 13pt

In the absence of transverse flow, a thermal source like (\ref{2.1}) 
depends on the particle rest mass and on the transverse momentum 
$K_\perp$ only through the combination $M_\perp^2 = m^2 +K_\perp^2$ 
(see Eq.~(\ref{2.5})). Furthermore, the source is then azimuthally and 
$x\to -x$ reflection symmetric. Hence $\langle \tilde x \tilde t 
\rangle$, $\langle \tilde x \tilde z \rangle$, and $\langle \tilde x^2 
- \tilde y^2\rangle$ all vanish and the approximations in 
Eqs.~(\ref{20b},\ref{20c}) become exact. As a result, all three YKP 
radii (\ref{20a})-(\ref{20c}) are only functions of $M_\perp$, too (as 
well as of Y, of course), i.e. they do not depend explicitly on the 
particle rest mass.  

\vspace*{12cm}
\includegraphics{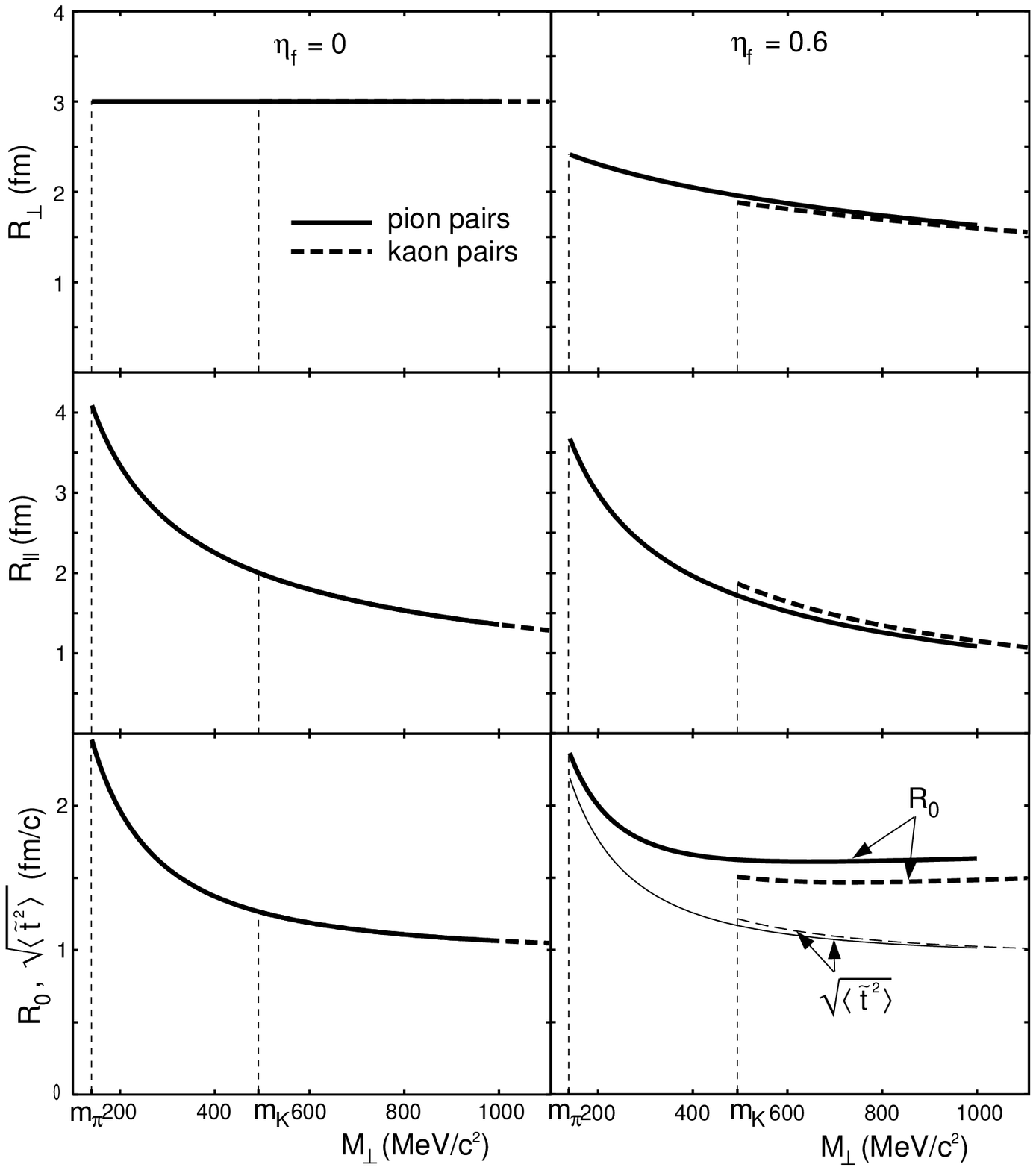}
\begin{center}
\begin{minipage}[t]{13cm}
\noindent {\footnotesize \bf Fig.5.  \rm
The YKP radii $R_\perp$, $R_\parallel$, and $R_0$ (from top to bottom)
for vanishing transverse flow (left column) and for $\eta_f=0.6$ (right
column), as functions of $M_\perp$ for pairs at $Y_{\rm cm}=0$. 
Solid (dashed) lines are for pions (kaons). The breaking of the
$M_\perp$-scaling by transverse flow is obvious in the right column.
Also, as shown in the lower right panel, for nonzero transverse flow 
$R_0$ does not agree exactly with the effective source lifetime 
$\protect\sqrt{\langle \tilde t^2\rangle}$. 
(Figure taken from Ref.~\protect\cite{WHTW96}.)}
\end{minipage}
\end{center}

This is seen in the left column of Fig.~5 where the three YKP radii 
are plotted for $Y_{\rm cm}=0$ pion and kaon pairs as functions of 
$M_\perp$; they agree perfectly. The transverse radius here shows no 
$M_\perp$-dependence due to the absence of transverse temperature 
gradients, but even with temperature gradients it would only depend on 
$M_\perp$. (Of course, this discussion neglects resonance decays 
which will be studied in Sec.~4.3.) The very strong 
$M_\perp$-dependence of the longitudinal radius parameter 
$R_\parallel$ is again due to the strong longitudinal expansion of the 
source. Note that $M_\perp$-scaling in the absence of transverse flow 
applies only to the YKP radius parameters: since the expressions 
(\ref{16b})-(\ref{16d}) involve nonvanishing variances with 
$\beta_\perp$- or $\beta_l$-prefactors (which depend explicitly on the 
rest mass), the HBT radii from the standard Cartesian fit do not 
exhibit $M_\perp$-scaling.  

For non-zero transverse flow $\eta_f\ne 0$ this $M_\perp$-scaling is 
broken by two effects: first, the second term in (\ref{2.5}) destroys
the $M_\perp$-scaling of the emission function itself, and second
the $\bbox{\beta}$-dependent correction terms in (\ref{20b},\ref{20c}) 
are now non-zero because the same term also breaks, for $K_\perp\ne 0$,
the $x \to -x$ and $x \to y$ symmetries. The magnitude of the associated
scale breaking due to the pion-kaon mass difference is seen in the right 
column of Fig.~5 for $\eta_f=0.6$. The effects are small and require very
accurate experiments for their detection. However, the sign of the effect
is opposite for $R_\parallel$ and for $R_\perp,\, R_0$ which may help 
to distinguish flow-induced effects from resonance decay contributions. 

Since for $Y_{\rm cm}=0$ the YK and CMS frames coincide, $\beta_l=0$ 
in the YK frame and the approximation in (\ref{20b}) remains exact 
even for non-zero transverse flow. The same is not true for the 
approximation in (\ref{20c}), and therefore I show in the lower right 
panel of Fig.~5 also the effective source lifetime $\sqrt{\langle 
\tilde t^2 \rangle}$ for comparison. The apparently rather large 
discrepancies between the YKP parameter $R_0$ and the effective source 
lifetime is due to a rather extreme choice of parameters: a large flow 
transverse flow and a small intrinsic source lifetime of $\Delta\tau = 
1$ fm/$c$ in (\ref{2.1a}). Since $\sqrt{\langle \tilde t^2 \rangle}$ 
approaches $\Delta\tau$ in the limit of large $M_\perp$ while the 
dominant \cite{WHTW96} correction term $\langle \tilde x^2 - \tilde 
y^2 \rangle$ does not depend on $\Delta\tau$, the YKP parameter $R_0$ 
will track the effective source lifetime more accurately for larger 
values of $\Delta\tau$ (and for smaller values of $\eta_f$).  

Why do $\sqrt{\langle \tilde t^2 \rangle}$ and $R_0$ increase at small 
$M_\perp$? Due to the rapid longitudinal expansion, the longitudinal 
region of homogeneity $R_\parallel$ is a decreasing function 
$M_\perp$. Since for different pair momenta $R_0$ measures the source 
lifetime in different YK reference frames, the freeze-out 
``hypersurface'' will in general appear to have different shapes for 
pairs with different momenta. Only in our model, where freeze-out 
occurs at fixed proper time $\tau_0$ (up to a Gaussian smearing with 
width $\Delta\tau$), is it frame-independent. It is thus generally 
unavoidable (and here, of course, true in any frame) that freeze-out 
at different points $z$ in the source will occur at different times 
$t$ in the YK frame. Since a $z$-region of size $R_\parallel$ 
contributes to the correlation function, $R_\parallel$ determines how 
large a domain of this freeze-out surface (and thus how large an 
interval of freeze-out times in the YK frame) is sampled by the 
correlator. This interval of freeze-out times combines with the 
intrinsic Gaussian width $\Delta\tau$ to yield the total effective 
duration of particle emission. It will be largest at small pair 
momenta where the homogeneity region $R_\parallel$ is biggest, and 
will reduce to just the variance of the Gaussian proper time 
distribution at large pair momenta where the longitudinal (and 
transverse) homogeneity regions shrink to zero. The rise of $\Delta 
t({\bf K})$ at small ${\bf K}$ is thus generic.  

While the strong $M_\perp$-dependence of the longitudinal radius 
parameter $R_\parallel$ arises from the strong longitudinal expansion, 
the weaker $M_\perp$-dependence of the transverse radius parameter 
reflects the weaker transverse expansion of our source. Following a 
suggestion by Th. Alber \cite{Alber95}, this relation can be made 
quantitative: in Fig.~6 we plot in the left column the transverse and 
longitudinal YKP radii $R_\perp$ and $R_\parallel$ versus $M_\perp$ on 
a double-logarithmic scale. We see that both can be approximately 
represented by power laws. (The same is not true for $R_0$.) While 
in such a plot the slope of $R_\perp$ clearly increases with the 
strength $\eta_f$ of the transverse flow, the slope of $R_\parallel$ 
appears to be insensitive to transverse flow. This can be seen 
quantitatively in the right column of Fig.~6 where we plot the powers
$\alpha_\perp,\alpha_\parallel$ extracted from a fit
 \begin{equation}
 \label{slope}
   R_\perp(M_\perp) \propto M_\perp^{-\alpha_\perp}\, ,
   \qquad
   R_\parallel(M_\perp) \propto M_\perp^{-\alpha_\parallel}
 \end{equation}
as a function of $\eta_f$. As indicated in Fig.~6b the extracted power 
$\alpha_\perp$ for $R_\perp$ depends somewhat on the fit region 
because $R_\perp$ doesn't follow an exact power law; independent of 
the fit region it increases, however, monotonously and nearly linearly 
with the strength $\eta_f$ of the transverse flow. Kaons ``feel" the 
transverse flow more strongly than pions, as reflected by the somewhat 
larger powers $\alpha_\perp$ at fixed $\eta_f$. Note that even for 
a rather strong transverse flow $\eta_f=0.6$ (heavy-ion data seem to 
require less flow) $\alpha_\perp$ remains below 0.25.

\vspace*{10cm}
\includegraphics{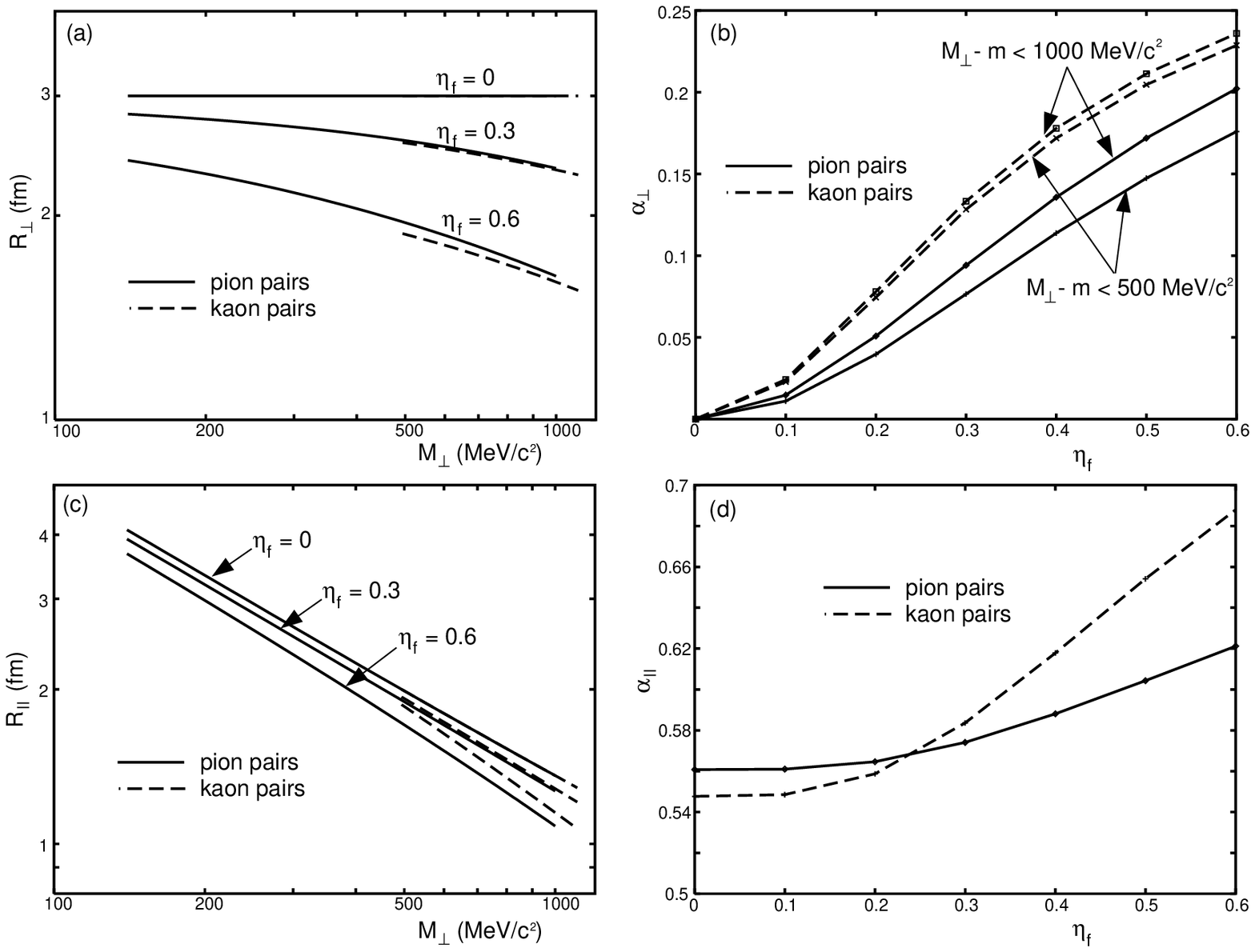}
\begin{center}
\begin{minipage}[t]{13cm}
\noindent {\footnotesize \bf Fig.6.  \rm
 (a) $R_{\perp}$ as a function of $\protect M_\perp$ at $Y_{_{\rm CM}}=0$,
     for pions (solid) and kaons (dashed) and 
     different transverse flow rapidities $\eta_f$.
 (b) The scaling coefficient $\alpha_{\perp}$ defined
     by $R_{\perp} \approx M_{\perp}^{-\alpha_\perp}$ for pions (solid)
     and kaons (dashed) as a function of the transverse flow rapidity 
     $\eta_f$. The different results obtained by fitting 
     in the regions $M_{\perp} - m < 500$ MeV/$c^2$ and 
      $M_{\perp} - m < 1000$ MeV/$c^2$ are shown separately.
 (c) Same as (a), but for $R_\parallel$.
 (d) Same as (b), but for $\alpha_\parallel$.
(Figure taken from Ref.~\protect\cite{WHTW96}.)}
\end{minipage}
\end{center}

The power $\alpha_\parallel$ for $R_\parallel$ is shown in Fig.~6d. It 
has a much larger value of 0.55-0.56 even for $\eta_f=0$, reflecting 
the strong boost-invariant longitudinal expansion. As $\eta_f$ 
increases, $\alpha_\parallel$ also increases, but only by a few 
percent. (Note the suppressed zero in Fig.~6d!) Again, kaons are 
affected more strongly by the transverse flow than pions, but 
altogether the $M_\perp$-dependence of $R_\parallel$ is rather 
insensitive to transverse flow.

As observed by Th. Alber in his thesis \cite{Alber95} these features
agree qualitatively with the heavy-ion data: for $\alpha_\parallel$
he found values of order 0.4--0.5, while $\alpha_\perp$ remained 
smaller, of order 0.1--0.2; in $^{32}$S-induced collisions, the 
central values for both numbers showed a systematic tendency to 
increase with the mass of the target nucleus, indicating stronger 
collective flow in larger collision systems than in smaller ones. The 
error bars on the $\alpha$'s were, however, large, and one should wait 
for independent confirmation before firmly drawing such a conclusion.

The $Y$-dependence of the Yano-Koonin rapidity $Y_{_{\rm YK}}$ and the 
$M_\perp$-dependence of $R_\parallel$ can thus be used as a 
quantitative measure for the longitudinal expansion of the source 
which is hardly affected at all by the presence and strength of 
transverse flow. On the other side, $R_\perp$ being boost-invariant, 
the $M_\perp$-dependence of $R_\perp$ is independent of the 
longitudinal expansion of the source and reflects only its transverse 
expansion. Together with the breaking of the $M_\perp$-scaling of the 
YKP radii, the $M_\perp$-dependence of $R_\perp$ can thus be used to 
extract quantitatively the transverse expansion velocity of the source 
\cite{HTWW96a,WHTW96}.

%
%
%

\vskip 13pt

\noindent{\bf 4.3.~Resonance decays}

\vskip 13pt

As already mentioned pions in particular have the problem that in 
high-energy collisions only a fraction of about 50\% or less of all 
pions come directly from the decoupling source while the rest are 
produced after decoupling by the decay of unstable resonances.
The above considerations presuppose that the resonance decays 
do not affect the $M_\perp$-dependence of $R_\perp$. This is not 
obvious, in particular since it is known that the $M_\perp$-dependence 
of the single-pion spectrum is very strongly affected by resonance 
decays \cite{Sollfrank}. Since resonance decays contribute more to 
pions than to kaons they may also affect the $M_\perp$-scaling 
arguments. The work by the Marburg group \cite{marb} on resonance 
decay effects on HBT in the context of hydrodynamical simulations 
indicates, within the standard Cartesian framework and without 
accounting for the cross-term, a possible additional 
$M_\perp$-dependence of the transverse radius. However, a systematic 
analysis of resonance contributions to HBT as a function of various 
characteristic source parameters is only now becoming available 
\cite{WH96,H96}.  

\vskip 13pt

\noindent{\it 4.3.1.~Formalism}

\vskip 13pt

Not much is known directly from experiment about the amount of 
resonance production in heavy-ion collisions. For $pp$ collisions at
similar energies a thermal model, where resonances are produced with 
thermal abundances in mutual chemical equilibrium, appears to work 
surprisingly well \cite{Hagedorn,Sollfrank,Becattini}. We therefore 
assume the same to hold for heavy-ion collisions. We also assume for 
simplicity that all hadrons decouple at the same point. Thus all 
resonances are assumed to have the same emission function 
(\ref{2.1}-\ref{2.1b}), adjusted only for the particle rest mass 
$m_i$, its spin $J_i$, and its chemical potential $\mu_i$. In chemical 
equilibrium $\mu_i$ is given in terms of the two independent chemical 
potentials $\mu_b$ and $\mu_s$ which account for conservation of 
baryon number and strangeness:
 \begin{equation}
 \label{chempot}
  \mu_i = b_i\, \mu_b + s_i \, \mu_s\, ,
 \end{equation}
where $b_i, s_i$ are the baryon number and strangeness for resonance
species $i$. $\mu_s$ is determined in terms of $T$ and $\mu_b$ by the 
condition of strangeness neutrality of the collision region 
\cite{LTHSR95} which cannot be violated on the time scale of strong 
interactions. For illustration we will below consider the case 
$\mu_b=\mu_s=0$. 

I will now describe very shortly the formal steps for calculating the 
resonance contributions to the correlation function; for more details 
I refer the reader to Ref.~\cite{WH96}. The total source distribution 
of pions (and similarly for other stable particle species, although 
for pions the resonance contributions are most important) can be 
written as 
 \begin{equation}
 \label{pionsource}
  S_\pi(x,p) = S_\pi^{\rm dir}(x,p) 
             + \sum_{r\ne \pi} S_{r\to\pi}(x,p)\, ,
 \end{equation}
where the first term on the r.h.s. is the contribution from the 
directly emitted pions (Eq.~(\ref{2.1}) with $J=0,m=m_\pi$) and the
sum contains all contributions from resonance feed-down:
 \begin{eqnarray}
 \label{Srpi}
  S_{r\to\pi}(x,p) &=& M_r \int_{s_-}^{s_+} ds \, g(s) 
       \int {d^3P\over E_{_{P}}}\, 
       \delta\left( p \cdot P - M_r E^* \right) 
 \nonumber\\
  &\times& \int d^4X \int_0^\infty d\tau\, \Gamma\, e^{-\Gamma\tau}\,
           \delta^{(4)} \left( x - \left( X-{P\over M_r} \tau \right) 
                        \right)\, S_r^{\rm dir}(X,P)\, .
 \end{eqnarray}
Here capital letters indicate coordinates associated with the parent 
resonance $r$, lower case letters are associated with the decay 
pion. $s$ is the the invariant mass of the other, unobserved decay 
products; in an $n$-body decay, it can vary between $s_- = (\sum_{i=2}^n 
m_i)^2$ and $s_+ = (M_r-m_\pi)^2$. $g(s)$ is the decay phase space for 
the $(n-1)$ unobserved particles; for the isotropic 2-body decay of an 
unpolarized resonance it is given by 
 \begin{equation}
 \label{2body}
   g(s) = {b_{r\to\pi} \over 4\pi p^*} \, \delta (s-m_2^2)\, ,
 \end{equation}
(where $b_{r\to\pi}$ is the branching ratio for the decay 
channel), and for isotropic 3-body decays by \cite{Hagedornbook}
 \begin{eqnarray}
 \label{3body}
  g(s) &=& {M_r b_{r\to\pi} \over 2\pi\, s}\,
       {\sqrt{[s-(m_2+m_3)^2][s-(m_2-m_3)^2]} \over 
        Q(M_r,m_\pi,m_2,m_3)}\, ,
 \\
  Q(M_r,m_\pi,m_2,m_3) &=& \int_{s_-}^{s_+} {ds'\over s'}
      \sqrt{(M_r+m_\pi)^2 - s'}
      \sqrt{s_+ - s'}
 \nonumber\\
  &&\qquad \times \ 
      \sqrt{s_- - s'}
      \sqrt{(m_2-m_3)^2 - s'} \, .
 \nonumber
 \end{eqnarray}
$p^*,E^*$ are the momentum and energy of the decay pion in the 
resonance rest frame,
 \begin{equation}
 \label{p*}
   E^* = \sqrt{m_\pi^2 + {p^*}^2}\, ,
   \qquad
   p^* = {\sqrt{[(M_r+m_\pi)^2 - s][(M_r-m_\pi)^2 - s]} \over
         2\, M_r}\, ,
 \end{equation}
and functions of $s$ only. The $\tau$-integration in (\ref{Srpi}) 
extends over the exponential decay probability of the resonance with 
total decay width $\Gamma$. The 4-dimensional $\delta$-function of the 
space-time coordinate $X$ ensures that for the pion to appear at point 
$x$ from a resonance decaying at time $\tau$, the parent resonance 
with momentum $P$ must have been emitted from the source at point 
$X-(P/M_r)\tau$. 

The integration over the resonance momentum $P$ is restricted by the 
energy-momentum constraint $\delta(p{\cdot}P - M_r E^*)$. In the 
coordinate system where the momentum $p$ of the decay pion is given by 
  \begin{equation} 
    \label{2.2a}
    p^\mu = (m_\perp\cosh y, p_\perp, 0, m_\perp\sinh y)\, ,
  \end{equation}
(see (\ref{2.2})), the resonance momentum $P$ is parametrized as
 \begin{equation}
 \label{Ppar}
    P^\mu = (M_\perp\cosh Y, P_\perp \cos\Phi, P_\perp \sin\Phi,
             M_\perp\sinh Y)\, .
  \end{equation}
For $p_\perp\ne 0$ the $\delta$-function can be used to fix the 
azimuthal angle $\Phi$ of the resonance momentum $P$ to
 \begin{equation}
 \label{3.7}
  \Phi_\pm = \pm \tilde \Phi \quad \text{with} \quad
  \cos \tilde \Phi = { E\, E_{_P} - p_{_L} P_{_L} - E^* M
                       \over p_\perp P_\perp}
                   = { m_\perp M_\perp \cosh(Y-y) - E^* M
                       \over p_\perp P_\perp}\, .
 \end{equation}
Let us denote by $P^\pm$ the two values of $P$ obtained by inserting 
the two solutions (\ref{3.7}) into (\ref{Ppar}). After doing the 
$\Phi$- and $X$-integrations in (\ref{Srpi}) one thus obtains
  \begin{eqnarray}
  \label{3.13}
    S_{r\to\pi}(x,p) 
    &=& M_r\, \int_{s_-}^{s_+} ds\, g(s)
        \int_{Y_-}^{Y_+} dY\, 
        \int_{M_{\perp,-}^2}^{M_{\perp,+}^2} dM_\perp^2 \,
        \int_0^{\infty} d\tau\, \Gamma e^{-\Gamma\tau} 
  \nonumber \\
    && \times\  
       { {1\over 2} \sum_\pm 
         S_r^{\rm dir} \left(x - {P^\pm\over M_r}\tau, P^\pm \right)
         \over 
         \sqrt{ p_\perp^2 (M_\perp^2 - M_r^2) -
                [E^* M_r - m_\perp M_\perp \cosh(Y-y)]^2 } }\, ,
 \end{eqnarray}
with the kinematic limits
  \begin{eqnarray}
  \label{3.11}
        M_{\perp,\pm} &=& \overline{M}_\perp \pm \Delta M_\perp
  \\
        &\equiv& { E^* M_r m_\perp \cosh(Y-y) \over
            m_\perp^2 \cosh^2(Y-y) - p_\perp^2 } 
          \pm 
          {M_r p_\perp 
           \sqrt{ {E^*}^2 + p_\perp^2 - m_\perp^2 \cosh^2(Y-y)}
           \over
           m_\perp^2 \cosh^2(Y-y) - p_\perp^2 }\,
  \nonumber\\ 
  \label{3.12}
        Y_\pm &=& y \pm \Delta Y \equiv 
        y \pm \ln \left( {p^* \over m_\perp} 
                       + \sqrt{ 1 + {{p^*}^2 \over m_\perp^2} } 
                  \right) 
  \end{eqnarray}
resulting from the zeroes of the square root in the denominator 
(which, incidentally, can also be written as $p_\perp P_\perp \vert 
\sin\tilde\Phi \vert$). -- For the limiting case $p_{\perp} = 0$, the 
constraint $p{\cdot}P = M_r E^*$ cannot be used to do the 
$\Phi$-integration. One then uses it to do the $M_{\perp}$-integral:
 \begin{eqnarray}
   S_{r\to\pi}(x;y,p_\perp=0) &=& 
        M_r\, \int_{s_-}^{s_+} ds\, g(s)
        \int_0^{\pi} d\Phi\,
        \int_{Y_-}^{Y_+} {\rm d}Y\,
        {M_r\, E^*\over m^2 \cosh^2(Y-y)} 
 \nonumber \\
   &&   \times \int d\tau \, \Gamma e^{-\Gamma\tau} \,
        S_r^{\rm dir} \left(x - {P\over M_r}\tau, P \right)
        \Bigg\vert_{M_\perp = {M_r E^* \over m_\pi\cosh(Y-y)}}\, .
 \label{3.10}
 \end{eqnarray}

For the more generic case $p_\perp\ne 0$ a few further manipulations 
are useful in practice \cite{WH96}: Rewriting the square root in 
(\ref{3.13}) as 
  \begin{equation}
  \label{3.14}
        {1\over \sqrt{m_\perp^2 \cosh^2(Y-y) - p_\perp^2}} \,
        {1\over \sqrt{(\Delta M_\perp)^2 - 
                      (M_\perp - \overline{M}_\perp)^2}}
  \end{equation}
and introducing new integration variables $v\in [-1,1]$, $\zeta \in [-
\pi,\pi]$ via 
  \begin{eqnarray}
  \label{3.15}
    M_\perp &=& \overline{M}_\perp + \Delta M_\perp \, \cos\zeta\, ,
  \\
  \label{3.16}
    Y &=& y  + v\, \Delta Y \, ,
  \end{eqnarray}
Eq.~(\ref{3.13}) can be further transformed into
  \begin{equation}
  \label{3.17}
        S_{R\to\pi}(x,p) = \sum_\pm \int_{\bf R} 
        \int_0^{\infty}{d\tau}\, \Gamma e^{-\Gamma\tau} 
        S_R^{\rm dir} \left( x -{P^\pm\over M} \tau,P^\pm \right) \, ,
 \end{equation}
with the following shorthand for the integration over the resonance 
momenta:
  \begin{equation}
  \label{3.18}
    \int_{\bf R} \equiv M_r \int_{s_-}^{s_+} ds\, g(s)
        \int_{-1}^1 {\Delta Y \, dv \over 
                   \sqrt{ m_\perp^2 \cosh^2 (v \Delta Y) - p_\perp^2} }
        \int_0^\pi d\zeta 
        \left( \overline{M}_\perp + \Delta M_\perp \cos\zeta \right) \, .
  \end{equation}

For the calculation of the correlation function we need the Fourier 
transform of the emission function
  \begin{equation}
  \label{3.19}
     \tilde S_{r\to\pi}(q,p) = 
     \int d^4x\, e^{iq{\cdot}x}\, S_{r\to\pi}(x,p) = 
     \sum_\pm \int_{\bf R} 
     {1 \over 1 - i { q{\cdot}P^\pm \over M_r\Gamma } } \,
     \tilde S_r^{\rm dir}(q,P^\pm) \, ,
  \end{equation}
and must evaluate
  \begin{equation}
  \label{3.23}
    C(\bbox{q},\bbox{K}) = 1 + 
    {  \vert\tilde S_{\pi}^{\rm dir}(q,K)\vert^2
     + 2\sum_{r\ne\pi} {\rm Re\,} [\tilde S_{\pi}^{\rm dir}(q,K)
                                   \tilde S_{r\to\pi}(q,K)]
     + \vert \sum_{r\ne\pi} \tilde S_{r\to\pi}(q,K)\vert^2 
     \over 
     \vert \tilde S_\pi(0,K) \vert^2} \, ,
  \end{equation} 
where the denominator includes all contributions. Numerically, this is 
a rather involved expression. If the resonance contributions are small 
compared to the direct term one can use the Grassberger 
approximation~\cite{G77} in which the last term in the numerator is 
neglected. For heavy-ion collisions this is not good enough since 
about 50\% of all pions come from resonance decays. Instead, we can 
try to exploit the connection from Lecture 2 between the half-widths 
of the correlation function and the space-time variances which are now 
given by
  \begin{equation}
    \langle \tilde x_\mu \tilde x_\nu \rangle(\bbox{K}) = 
    {\sum_r \int d^4x\, \tilde x_\mu \tilde x_\nu \, S_{r\to\pi}(x,K) 
     \over
     \sum_r \int d^4x\, S_{r\to\pi}(x,K)} \, .
    \label{3.24}
  \end{equation}
Here the sum now runs over all contributions, including the direct 
pions. It is instructive to rewrite the average over the emission 
function in the following form:
  \begin{eqnarray}
  \label{3.25}
    \langle x_\nu \rangle (\bbox{K})  
    &=& \sum_r f_r(\bbox{K}) \ 
        \langle x_\nu \rangle_r (\bbox{K})\, , 
  \nonumber\\
    \langle x_\mu x_\nu \rangle (\bbox{K})  
    &=& \sum_r f_r(\bbox{K}) \ 
        \langle x_\mu x_\nu \rangle_r (\bbox{K})\, .
  \end{eqnarray}
Here we introduced the single-particle fractions \cite{marb}
  \begin{equation}
    f_r(\bbox{K}) = {\int d^4x\, S_{r\to\pi}(x,K) \over
                     \sum_r \int d^4x\, S_{r\to\pi}(x,K)} 
                   = {dN_\pi^r/d^3K \over dN_\pi^{\rm tot}}\, ,
    \quad
    \sum_r f_r(\bbox{K}) = 1\, ,
  \end{equation}
which give the fraction of single pions with momentum $\bbox{K}$ 
resulting from decay channel $r$, and the average $\langle \dots 
\rangle_r$ with the effective pion emission function arising from
this particular channel: 
  \begin{equation}
  \label{3.26}
    \langle \dots \rangle_r(\bbox{K}) = 
    {\int d^4x \dots S_{r\to\pi}(x,K) \over
     \int d^4x\, S_{r\to\pi}(x,K)} \, .
  \end{equation}
The variances (\ref{3.24}) can then be rewritten as
  \begin{equation}
    \langle \tilde x_\mu \tilde x_\nu \rangle 
    = \sum_r f_r\, \langle \tilde x_\mu \tilde x_\nu \rangle_r 
    + \sum_{r,r'} f_r (\delta_{r,r'} - f_{r'}) 
    \langle x_\mu \rangle_r \langle x_\nu \rangle_{r'}\, . 
  \label{3.24a}
  \end{equation}
The first term has an easy intuitive interpretation: each resonance 
decay channel $r$ contributes an effective emission function 
$S_{r\to\pi}$. The full variance is calculated by weighting the 
variance (homogeneity length) of the emission function from a 
particular decay channel with the fraction $f_r$ with which this 
channel contributes to the single particle spectrum. The second term 
in (\ref{3.24a}) is due to the fact that in general the effective
emission functions from the various decay channels have different 
saddle points; it somewhat spoils the intuitive interpretation of 
(\ref{3.24}).

\vskip 13pt

\noindent{\it 4.3.2.~Influence on HBT radii and non-Gaussian features}

\vskip 13pt

It turns out that, contrary to the situations discussed before, in the 
case of long-lived resonances the expressions (\ref{3.24}) are not 
very useful for a quantitative understanding of the correlator, 
although certain qualitative features can still be extracted 
relatively easily. The reason for this is best explained by 
considering the simple example of only one longlived resonance in a 
1-dimensional space. Let us model the emission function for the 
direct pions by a Gaussian in coordinate space with width $R_{\rm 
dir}$ and (somewhat unrealistically) the emission function of the 
pions from the decaying resonance by a second Gaussian with much 
larger radius $R_{\rm halo}$ (assuming that the resonance travels on 
average a distance of order $R_{\rm halo}$ before it decays), with 
weights $\varepsilon$ and $(1-\varepsilon)$, respectively:
 \begin{equation}
 \label{halo}
   S_\pi(x,K) = S_\pi^{\rm dir}(x,K) + S_{r\to\pi}(x,K) =
   (1-\varepsilon)\, e^{-x^2/(2R_{\rm dir}^2)} + 
   \varepsilon\, e^{-x^2/(2R_{\rm halo}^2)}\, .
 \end{equation}
Then the correlator is given by
 \begin{equation}
 \label{halo1}
   C(q,K) - 1 = (1-\varepsilon)^2\, e^{-R_{\rm dir}^2 q^2} + 
   \varepsilon^2\, e^{-R_{\rm halo}^2 q^2} +
   2 \varepsilon (1-\varepsilon) \,
   e^{-(R_{\rm dir}^2 + R_{\rm halo}^2) q^2/2}\, .
 \end{equation}
If $\varepsilon$ is small, but $R_{\rm halo}$ is large, then
the correlator is a superposition of a large, broad Gaussian with 
width $1/R_{\rm dir}$ and weight $(1-\varepsilon)^2$, a second, 
narrower Gaussian with width $\sqrt{2/(R_{\rm dir}^2 + R_{\rm 
halo}^2)}$ and smaller weight $2\varepsilon(1-\varepsilon)$, and a 
third, extremely narrow Gaussian with width $1/R_{\rm halo}$ and tiny 
weight $\varepsilon^2$. Obviously, the rough structure of the 
correlator will be determined by the large and broad direct 
contribution; the two other contributions will, however, modify its 
functional form: 
\newline
(i) If the resonance is shortlived such that $R_{\rm halo} \gapp 
R_{\rm dir}$ its effect on the correlator will be minor; its shape 
will remain roughly Gaussian, with a width somewhere between $1/R_{\rm 
dir}$ and $1/R_{\rm halo}$, depending on the weight $\varepsilon$ of 
the resonance contribution.  
\newline
(ii) If the resonance lifetime and thus $R_{\rm halo}$ are extremely 
large, the second and third term in (\ref{halo1}) will be very narrow 
and, due to the finite two-track resolution of every experiment, may 
escape detection; then the correlator looks again Gaussian with a 
width $1/R_{\rm dir}$, but at $q=0$ it will not approach the value 2, 
but $1 + (1-\varepsilon)^2 < 2$. The correlation appears to be 
incomplete, with an ``chaoticity parameter" $\lambda = (1-f_r)^2 
= (1-\varepsilon)^2$. 
\newline
(iii) If the resonance lifetime is in between such 
that $R_{\rm halo} \gg R_{\rm dir}$ but $1/R_{\rm halo}$ being still 
large enough to be experimentally resolved, all three Gaussians 
contribute, and the full correlator deviates strongly from a single 
Gaussian.  

In cases (ii) and (iii) the space-time variances give misleading or 
outright wrong results for the width of the correlation function. As noted
in connection with Eq.~(\ref{corrgauss}), they reproduce the curvature
of the correlator at $q=0$ which for our toy model is
 \begin{equation}
 \label{curv}
   {1\over 2}
   \left. {\partial^2 C(q)\over \partial q^2} \right\vert_{q=0} = 
   (1-\varepsilon) R_{\rm dir}^2 + \varepsilon R_{\rm halo}^2 \, .
 \end{equation}
In case (ii), for not too small values of $\varepsilon$, this is 
dominated by the second term although the resonance contribution is 
not even visible in the correlator! On a quantitative level, the 
situation is not very much better for case (iii). 

For the case (ii) of very long-lived resonances there is, of course, an 
easy way to save the usefulness of the space-time variances: if one 
simply leaves them out from the sum over resonances in (\ref{3.24a}),
but only includes them via an ``chaoticity parameter" 
 \begin{equation}
 \label{lambda}
   \lambda(\bbox{K}) = \left( 
   1 - \sum_{r = {\rm longlived}} f_r(\bbox{K}) \right)^2\, , 
 \end{equation}
the roughly Gaussian contributions to the correlator from the direct 
pions and shortlived resonances are still correctly reproduced. The 
real head ache comes from resonances with an intermediate lifetime 
which lead to a large halo but can still be experi\-mentally resolved. 
They cause appreciable deviations from a Gaussian behaviour for the 
correlator and cannot be reliably treated by the method of space-time 
variances. 

In nature there is only one such resonance: the $\omega$ meson, with a 
lifetime of approximately 20 fm/$c$. All other resonances either live 
so shortly (typically 1 fm/$c$) that they hardly modify the 
correlator, or so long that their contribution to the correlator 
cannot be resolved such that they only affect $\lambda$. At low 
$K_\perp$, however, up to 10\% of all pions come from $\omega$-decays
($f_\omega(\bbox{K}=0) \approx 0.1$), and their non-Gaussian effects 
on the correlator can be clearly seen. On account of the $\omega$ a 
full numerical evaluation of the correlator (\ref{3.23}) or a 
treatment with more powerful analytical methods which can deal with 
the non-Gaussian features of the correlator ($q$-variances, see 
Ref.~\cite{WH96}) become indispensible.

\vskip 13pt

\noindent{\it 4.3.3.~$K$-dependence of correlator including resonance 
decays}

\vskip 13pt
 
I now show some numerical results for the correlation functions 
resulting from the emission function (\ref{2.1}), but now including 
the resonance contributions. The complete spectrum of relevant 
resonances is included, and in the decays the 2- and 3-body decay 
kinematics is fully taken into account. The HBT radii are extracted 
from a Gaussian fit to the numerically calculated correlation 
function. A detailed technical discussion is given in 
Ref.~\cite{WH96}.  

\vspace*{10cm}
\includegraphics{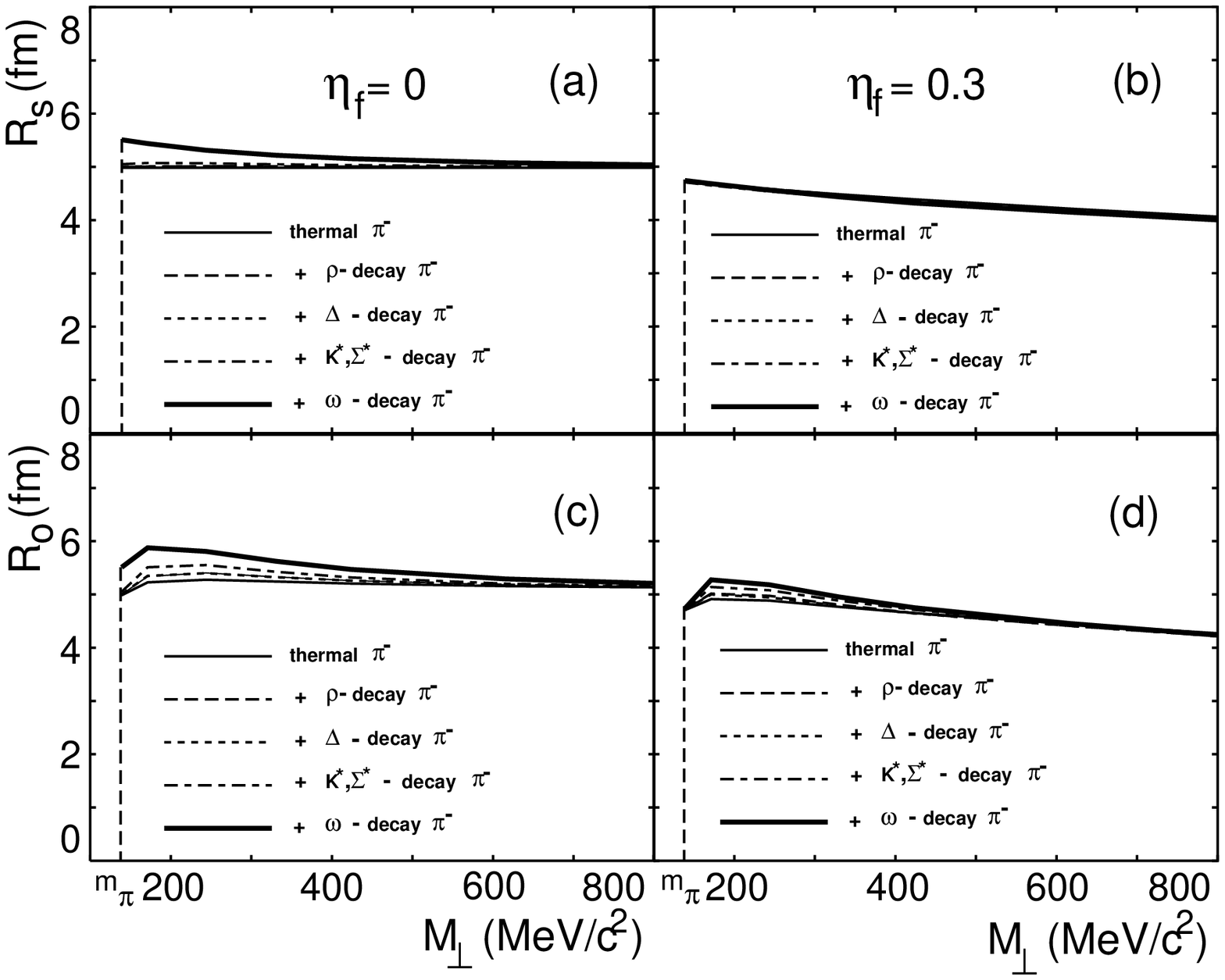}
\begin{center}
\begin{minipage}[t]{13cm}
\noindent {\footnotesize \bf Fig.7.  \rm
The influence of resonance decays on the $M_\perp$-dependence of
$R_s$ (a,b) and $R_o$ (c,d) for $Y_{\rm cm}=0$ pion pairs. a,c: no 
transverse flow; b,d: transverse flow rapidity $\eta_f=0.3$.
The Gaussian transverse radius is here $R=5$ fm, and $T=150$ MeV.
The HBT radii are extracted from unidirectional fits to the correlator 
in the respective direction of $\bbox{q}$.
(Figure taken from Ref.~\protect\cite{WH96}.)}
\end{minipage}
\end{center}

Fig.~7 shows results for the standard Cartesian parameters $R_s$ and 
$R_o$ from 1-dimensional fits to the numerically computed correlator 
in the respective ($q_s$ or $q_o$) directions (setting the other 
components of $\bbox{q}$ to zero). One sees that the effects of the 
short-lived resonances with lifetimes of order 1 fm/$c$ on $R_s$ are 
essentially negligible, both at vanishing and at nonzero transverse 
flow. Only the $\omega$ with its intermediate lifetime of 20 fm/$c$ 
affects $R_s$, but only for vanishing transverse flow. There it 
induces a weak $M_\perp$-dependence at small $M_\perp$ even in the 
absence of transverse flow; at $M_\perp>500$ MeV the contribution of 
the $\omega$ dies out, and $R_s$ again becomes $M_\perp$-independent 
(which would not be the case if it were affected by flow). At 
$\eta_f=0.3$ and 0.6 \cite{WH96} not even the $\omega$ generates any 
additional $M_\perp$-dependence! --$R_o$ shows some effects from the 
additional lifetime of the resonances, in particular from the 
long-lived $\omega$. Resonances with much longer lifetimes than the 
$\omega$ (in particular all weak decays) cannot be resolved
experimentally in the correlator and have no effect on the HBT radii.

\vspace*{16cm}
\includegraphics{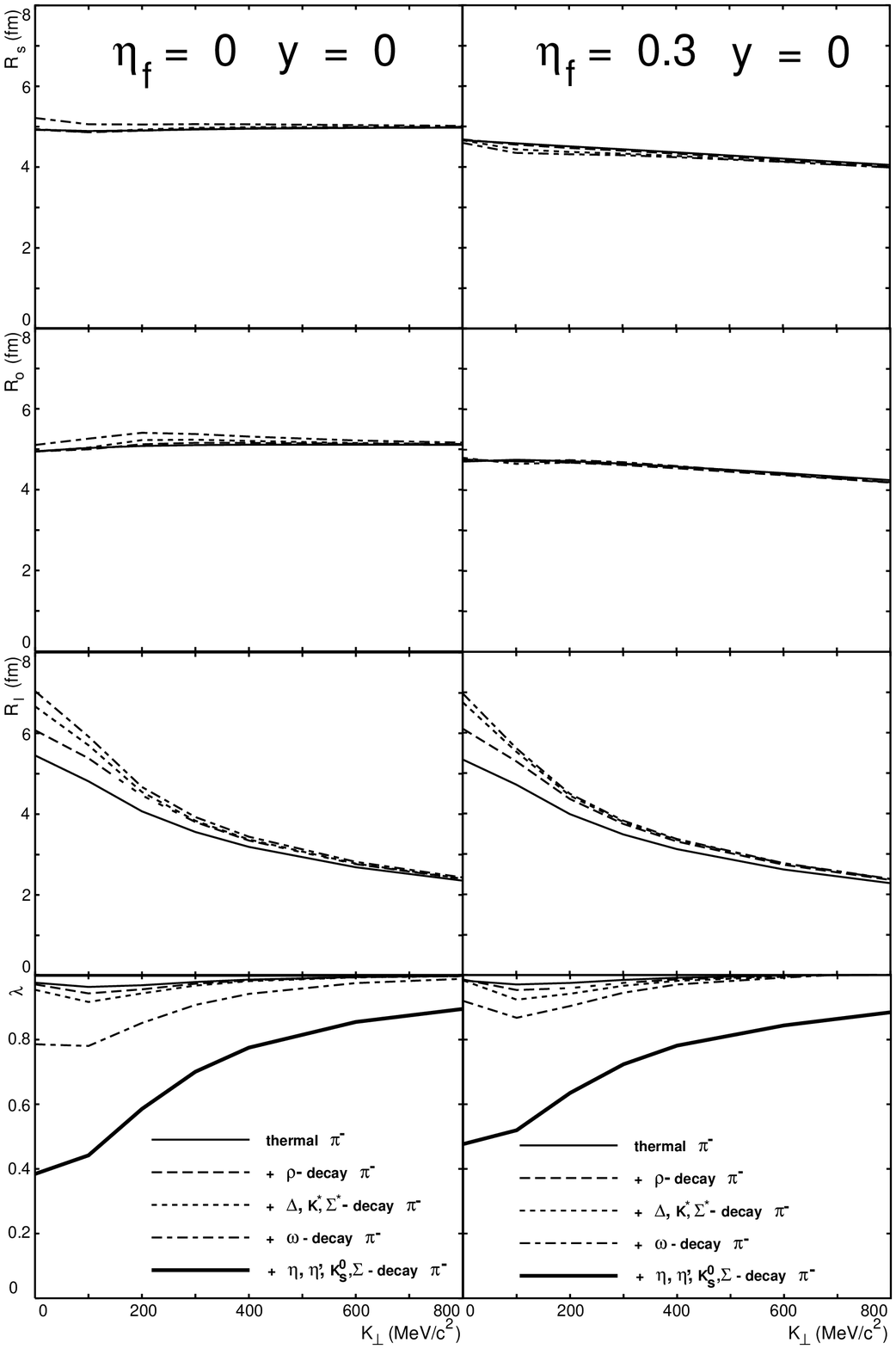}
\begin{center}
\begin{minipage}[t]{13cm}
\noindent {\footnotesize \bf Fig.8.  \rm
Same as Fig.~7, but now with parameters extracted from a complete 
5-dimensional fit to the correlator (see text). Top row: $R_s$;
second row: $R_o$; third row: $R_l$; bottom row: effective chaoticity 
parameter $\lambda$. Left column: no transverse flow; right column:
transverse expansion with $\eta_f=0.3$. The cross term $R_{ol}$ 
vanishes for $Y=0$ pairs.
(Figure taken from Ref.~\protect\cite{WH96}.)}
\end{minipage}
\end{center}

The weak effect of resonances on $R_s=R_\perp$ seems surprising: due 
to their non-zero lifetime they should be able to propagate outside the 
original source before decay and form a pion ``halo'' \cite{marb,CLZ94}.
This effect is, however, much weaker than naively expected: most
of the resonances are not very fast, and the halo thickness is thus only
a fraction of the resonance lifetime. At finite transverse flow an
additional effect comes into play: it turns out that then the effective 
size of the emission function for directly emitted resonances is 
{\em smaller} than that for direct pions \cite{WH96}! At $\eta_f{=}0.3$ 
and 0.6 this even slightly overcompensates the halo effect, and altogether 
the resonances change neither the size nor the $M_\perp$-dependence 
of $R_s$. 

In Fig.~8 also the longitudinal radius $R_l$ is shown. Here even the 
shortlived resonances are seen to make an effect. It can be 
essentially traced to a lifetime effect: Since the pions from the
short-lived resonances appear typically 1 fm/$c$ later than the direct 
pions, the source has in the meantime expanded longitudinally to a 
situation with a smaller longitudinal velocity gradient (the latter 
goes like $1/\tau$). Thus the resulting pion source has a larger 
longitudinal length of homogeneity and features a larger value $R_l$.  

\vskip 13pt

\noindent{\it 4.3.4.~The ``chaoticity parameter" $\lambda$}

\vskip 13pt

As noted above, the $\omega$-decays make the correlator non-Gaussian.
Slight non-Gaussian features exist even without resonance decays
in the longitudinal direction, induced by the strong longitudinal 
flow, and for large transverse flow $\eta_f$ become also visible in 
the transverse direction. A Gaussian fit to such a slightly non-Gaussian 
correlator in general does not extrapolate to the correct value at 
$\bbox{q}=0$, but introduces an effective ``chaoticity parameter" 
$\lambda$. In 1-dimensional fits along the three cartesian directions 
of $\bbox{q}$ the different degrees of non-Gaussicity lead to 
different values of $\lambda$. This situation is aggravated by the 
resonance contributions, in particular from the $\omega$ which affects
the correlator differently in each direction. Still a different value 
of $\lambda$ is found in a 5-dimensional fit to the correlator, using 
the 4 standard Cartesian HBT radii and $\lambda$ as fit parameters, 
because now $\lambda$ has to ``compromise" between the values found
in the unidirectional fits. From the last row in Figs.~8 one sees that 
even without the contribution from the $\omega$ and the longlived 
resonances the effective $\lambda$ is below 1 by up to 10\%; inclusion 
of the $\omega$ reduces it further to about 80-85\% at $\bbox{K}=0$.
Thus not only the very longlived resonances affect $\lambda$ as 
anticipated above, but so do to some extend the medium- and shortlived 
resonances and even flow.
 
Comparing Figs.~7 and Fig.~8 one sees that this compromise in 
$\lambda$ between the 3 different unidirectional fit values and the 
one resulting in the 5-dimensional fit also changes the fitted HBT 
radii. On the same (small) level as already observed for the
contribution from $\omega$ decays, it even affects the 
$M_\perp$-dependence of $R_\perp$. Both effects, the non-Gaussian 
features of the correlator and the effective ``chaoticity parameter"
which differs from $\lambda =1$, vanish at large $\bbox{K}$ because
the resonance decay pions are concentrated at low $\bbox{K}$ 
\cite{Sollfrank}.

\vskip 26pt

\noindent{\bf 5.~FINAL REMARKS}

\vskip 13pt

In these three lectures I have presented to you two-particle intensity 
interferometry for relativistic heavy-ion collisions both as an 
intellectually stimulating problem and as a powerful practical method.
Its application to nuclear collisions has turned out to be much more 
difficult than expected from the astrophysical analogue (two-photon 
intensity interferometry of stars). But at the same time, due to the 
dynamical nature of the problem, the physics of heavy-ion collisions 
is very much richer, and I have tried to show you that the HBT method 
is up to the task of clarifying a lot of this physics in a rather 
direct manner.  

The key to our understanding of the HBT method and how to apply it to 
dynamical situations are the model-independent expressions derived in 
the second lecture, which express the HBT width parameters in terms of 
second order space-time variances of the emission function. They 
provide the basis of a detailed physical interpretation of the 
measured HBT radii. They show that generally the HBT radius parameters 
do not measure the full geometric extension of the source, but regions 
of homogeneity inside the effective emission function for particles with 
certain fixed momenta. For expanding systems these are usually 
smaller than the naive geometric source size and decreasing functions 
of the pair momentum. For systems with finite lifetime the HBT 
parameters usually mix the spatial and temporal structure of the 
source, and their unfolding requires model studies.  

With the new YKP parametrization a method has been found which, for 
systems with dominant longitudinal expansion, cleanly factorises the 
longitudinal and transverse spatial from the temporal homogeneity 
length. The effective source lifetime is directly fitted by the 
parameter $R_0$; it is generically a function of the pair momentum and 
largest for pairs which are slow in the CMS. Another fit parameter, 
the YK velocity, measures directly the longitudinal velocity of the 
emitting fluid element, and its dependence on the pair rapidity allows 
for a direct determination of the longitudinal expansion of the 
source. Without transverse expansion, the YKP radius parameters show 
exact $M_\perp$-scaling. The breaking of this scaling and the 
$M_\perp$-dependence of the transverse radius parameter $R_\perp$ 
allow for a determination of the transverse expansion velocity of the 
source. Resonance decays were shown to mostly affect the lifetime 
parameter and, as a consequence, the longitudinal homogeneity length.
They leave the $M_\perp$-dependence of $R_\perp$ nearly unchanged and  
thus do not endanger the extraction of the transverse flow via HBT.  

\vskip 13pt

With this new and detailed understanding of the method, I believe 
that HBT interferometry has a begun a new and vigorous life as a 
powerful tool for reconstructing the geometric and dynamic space-time 
characteristics of the collision zone from the measured momentum spectra. 

\vskip 13pt

\noindent {\bf Acknowledgements:} 
I wish to express my thanks to my collaborators on this project, 
S.~Chapman, J.R. Nix, B. Tom\'a\v sik, U.A. Wiedemann, and Wu Yuanfang, 
who each contributed valuable pieces to the results presented in these 
lectures. Working with them has always been a pleasure. I would also 
like to acknowledge stimulating discussions with H. Appelsh\"auser, 
T. Cs\"org\H o, D. Ferenc, M. Ga\'zdzicki, B. Jacak, and P. Seyboth. 
Last but not least I would like to thank the organizers of this summer 
school for the invitation to come here and the students for their 
enthusiasm and for a pleasant and entertaining week in Dronten with 
many interesting conversations. This work was supported by grants from 
BMBF, DFG, and GSI.

%
%
%


%
%

\newpage

\noindent
{\bf Keywords:} Two-particle interferometry / Hanbury-Brown-Twiss 
interferometry / HBT radii / relativistic heavy-ion collisions / quark-gluon
plasma / resonance decays / collective flow / hydrodynamics / YKP 
parametrization / Mperp-scaling / emission function

\noindent 
{\bf Abstract:}
I discuss two-particle intensity interferometry as a method to extract from 
measured 1- and 2-particle momentum spectra information on the space-time 
geometry and dynamics of the particle emitting source. Particular attention 
is given to the rapid expansion and short lifetime of the sources created
in relativistic heavy-ion collisions. Model-independent expressions for 
the HBT size parameters in terms of the space-time variances of the source 
are derived, and a new parametrization of the correlation function is 
suggested which allows to separate the transverse, longitudinal and 
temporal extension of the source and to measure its transverse and 
longitudinal expansion velocity. The effects of resonance decays are also
discussed.

\end{document}